\documentclass[a4paper,fleqn]{cas-dc}
\usepackage{cas-dfrws} %overwrite some of the original (cas-dc.cls)
\renewcommand{\dfrwsconference}{Digital Forensics Doctoral Symposium (DFDS),  April 1, 2025}

\usepackage{amssymb}

\usepackage[normalem]{ulem}
\usepackage{booktabs}
\usepackage{array}
\usepackage{comment}
\newcolumntype{x}[1]{>{\centering\arraybackslash\hspace{0pt}}p{#1}}

\definecolor{ao}{rgb}{0.0, 0.5, 0.0}

\newcommand{\black}{\color{black}}

\definecolor{blueCV}{RGB}{204,229,255}
\definecolor{orangeCV}{RGB}{255,169,179}
\definecolor{redCV}{RGB}{255,220,150}

\usepackage{subcaption}

\usepackage[authoryear]{natbib}
\usepackage{tikz}
\usetikzlibrary{backgrounds,fit, positioning, arrows, arrows.meta, calc, decorations.pathreplacing,calligraphy}
\usepackage{pgfplots}
\pgfplotsset{compat=1.18}
\usepackage{pgfplotstable}
\usepackage{subcaption}

\PassOptionsToPackage{hyphens}{url}
\hypersetup{
    colorlinks,
    linkcolor=blue,
    citecolor=blue,
    urlcolor=blue
}

\expandafter\def\expandafter\UrlBreaks\expandafter{\UrlBreaks% save the current one 
    \do\a\do\b\do\c\do\d\do\e\do\f\do\g\do\h\do\i\do\j% 
    \do\k\do\l\do\m\do\n\do\o\do\p\do\q\do\r\do\s\do\t% 
    \do\u\do\v\do\w\do\x\do\y\do\z\do\A\do\B\do\C\do\D% 
    \do\E\do\F\do\G\do\H\do\I\do\J\do\K\do\L\do\M\do\N% 
    \do\O\do\P\do\Q\do\R\do\S\do\T\do\U\do\V\do\W\do\X% 
    \do\Y\do\Z\do\*\do\-\do\~\do\'\do\"\do\-}% 
\usepackage[hyphenbreaks]{breakurl}

\usepackage[capitalise]{cleveref}
\usepackage{enumitem}

\newcommand{\urlDate}{last accessed 2021-02-22}
\newcommand{\furl}[1]{\footnote{\url{#1} (\urlDate).}}

\usepackage{version}
\excludeversion{finalversion}
\excludeversion{detailedversion}
% ...
\begin{document}

%\title[mode=title]{Understanding Time Bandits: A User Study on Live Temporal Tampering and Its Effect on Event Reconstruction}
%\shorttitle{Tripping over Time}

% Back to the Future: A user study to understand how they manipulate and tamper wth timestamps

% OUTATIME: ... (a more niche back to the future reference)

% Time Warp: ...

% Lets find a time warp again: ... (full on 1970 Rocky Horror Show reference)

\title[mode=title]{Strategies and Challenges of Timestamp Tampering for Improved Digital Forensic Event Reconstruction (extended version)}
\tnotemark[1]
\tnotetext[1]{This is the extended version of the article accepted at Digital Forensics Doctoral Symposium part of DFRWS EU 2025.}

%``Tamperer, thou shall not succeed'': Learnings from a user study on tampering for improved event reconstruction}
\shorttitle{Strategies and Challenges of Timestamp Tampering}
\shortauthors{C.~Vanini et al.}

%\title{Event reconstruction and tampering evidence -- An experiment on what to trust}
% Title ideas for brainstorming:
%
% Tripping over time: a user study on temporal tampering for improved digital forensic event reconstruction
% Swap it like it's hot: a user study on the manipulation of event order 
% Time is relative: a user study on timestamp manipulations
% Grain of time: insights from tampering observations
% T-1: a user study on temporal tampering
% Resist the tamperer: insights from tampering observations
% Tamperer, thou shall not succeed: Learnings from a user study on tampering 

\author[1]{Céline Vanini}[orcid=0000-0003-3690-8574]
\credit{Conceptualization, Methodology, Investigation,
  Writing - Original draft, Writing - Review \& Editing}
\ead{celine.vanini@unil.ch}

\author[3]{Jan Gruber}[orcid=0000-0003-1862-2900]
\ead{jan.gruber@fau.de}
\credit{Conceptualization, Writing - Original draft, Writing - Review \& Editing, Visualization}

\author[2]{Christopher J. Hargreaves}
\credit{Conceptualization, Methodology, Supervision, Writing - Review \& Editing}
\ead{christopher.hargreaves@cs.ox.ac.uk}

\author[3]{Zinaida Benenson}[orcid=0009-0006-7158-0219]
\ead{zinaida.benenson@fau.de}
\credit{Methodology, Supervision, Writing - Review \& Editing}

\author[3]{Felix Freiling}[orcid=0000-0002-6128-5201]
\ead{felix.freiling@fau.de}
\credit{Conceptualization, Funding Acquisition, Methodology, Supervision, Writing - Review \& Editing}

\author[1,4]{Frank Breitinger\corref{*}}[orcid=0000-0001-5261-4600]
\ead{frank.breitinger@uni-a.de}
\ead[url]{https://FBreitinger.de}
\credit{Conceptualization, Methodology, Supervision, Writing - Review \& Editing}

\affiliation[1]{organization={School of Criminal Justice},
             addressline={University of Lausanne},
             %city={},
             %postcode={1015},
             state={Lausanne},
             country={Switzerland}}

\affiliation[2]{organization={Department of Computer Science},
             addressline={University of Oxford},
             %city={},
             %postcode={},
             %state={Oxford, OX1 3QD},
             country={United Kingdom}}

\affiliation[3]{organization={Department of Computer Science},
             addressline={Friedrich-Alexander Universität (FAU)},
             %city={},
             %postcode={2497},
             state={Erlangen},
             country={Germany}}

\affiliation[4]{organization={Present affiliation: Institute of Computer Science},
             addressline={University of Augsburg},
             %city={},
             %postcode={2497},
             state={Augsburg},
             country={Germany}}

\begin{abstract}
Timestamps play a pivotal role in digital forensic event reconstruction, but due to their non-essential nature, tampering or manipulation of timestamps is possible by users in multiple ways, even on running systems. This has a significant effect on the reliability of the results from applying a timeline analysis as part of an investigation. 
In this paper, we investigate the problem of users tampering with timestamps on a running (``live'') system. While prior work has shown that digital evidence tampering is hard, we focus on the question of \emph{why} this is so. By performing a qualitative user study with advanced university students, we observe, for example, a commonly applied multi-step approach in order to deal with second-order traces (traces of traces). %We also identify factors that allow the derivation of investigative leads to detect tampering. For example, study participants generally concentrated on artifacts that can be manipulated more easily than others. 
We also derive factors that influence the reliability of successful tampering, such as the individual knowledge about temporal traces, and technical restrictions to change them. These insights help to assess the reliability of timestamps from individual artifacts that are relied on for event reconstruction and subsequently reduce the risk of incorrect event reconstruction during investigations. 
\end{abstract}


\maketitle

\section{Introduction}
\label{introduciton}

%\emph{``Pleading not guilty!''}---Criminals have a vital interest in concealing the crime they committed, their connection to it, or their identity to evade the pressure of prosecution~\citep{Turvey22}. Therefore, perpetrators often employ evidence tampering to create fictitious or fake traces to sabotage an investigation. Tampering, as part of general evidence dynamics~\citep{ChisumT00}, is commonly considered to be easier when dealing with digital evidence~\citep{KumarSL05,Caloyannides03,Lin18}; hence, it is more likely that the perpetrators try to remove or modify digital traces they have left behind when committing their crimes. This is particularly true for timing information because it is critical to establishing when and in which order certain actions have happened.

The dangers of evidence tampering, i.e., the intentional act of altering, concealing or falsifying evidence \citep{Sanchirico:2004:ET}, are of great concern to law enforcement agencies since fictitious or fake traces can easily alter or sabotage a criminal investigation \citep{ChisumT00}. While there is a high awareness of the risks of tampering for physical evidence, tampering of digital evidence is much less understood. While some scholars consider tampering to be easier when dealing with digital evidence~\citep{Caloyannides03,Lin18}, others claim that it is at least similarly difficult in special cases \citep{DBLP:journals/di/SchneiderWF20}. Understanding the risks of evidence tampering is particularly important for \emph{timestamps} because, firstly, timestamps play a pivotal role in digital forensic event reconstruction to establish the order in which certain actions happened, and secondly, timestamp manipulation is a commonly applied indicator removal technique in security incidents \citep{timestoming, StromAMNPT20}.

\subsection{Related Work}

Despite work that has structured tampering activities under the heading of anti-forensics \citep{Harris06, Garfinkel07, conlan_anti-forensics_2016}, and confirmed by the literature review of \citet{neale_fool_2023}, unfortunately, our understanding of digital evidence tampering in general and of timestamp tampering in particular is rather shallow. Previous research has primarily focused on specific technical contexts of timestamp tampering, such as NTFS \citep{galhuber_time_2021, mohamed_detection_2019, palmbach_artifacts_2020}, or on technical approaches for timestamp tampering detection. These basically attempt to find inconsistencies between timestamps, e.g., the violation of general time rules \citep{galhuber_time_2021}, of causal relationships between timestamps \citep{Willassen2008, MARRINGTON2011S52} or inconsistent relations to implicit timing information like sequence numbers \citep{Dreier:2024:BT}. Even if such inconsistencies are detected, it may still be unclear whether these are due to intentional tampering. For example, if some timestamps have been set to January 1, 1970, explanations can vary from intentional ``timestomping'' \citep{timestoming} to an unintentional non-initialized Unix timestamp \citep{Manjoo:2001:UTT}. Thus, it is still necessary to have a solid understanding of adversarial (tampering) behaviors such that investigators can assess the effect of tampering on the meaning of evidence. 

Some previous user studies \citep{Moch15, freiling_controlled_2018, DBLP:journals/di/SchneiderWF20} also focused on the tampering detection problem. In these tampering detection experiments, participants had to produce convincing forgeries that should be taken as originals when analyzed by other participants in the study. This allowed a confusion matrix to be created and false positive rates calculated. 
%Due to their empirical study design and the low number of participants, these experiments did not establish the statistical significance of their results, nor did they attempt to investigate tampering strategies employed by study participants. 
Due to their empirical study design, these experiments did not attempt to investigate the difficulties of tampering with specific artifacts.
Furthermore, the experiment setup considered the extreme case where a perpetrator has \emph{full control} over every bit of the system, an approach we call \emph{dead tampering}. Still, the quantitative insights from these works indicate that tampering may not be as easy as it can be expected, but it is highly unclear \emph{why} this may be so. 
%This research question, together with the commonly observed shortage of skilled study participants, therefore calls for a more \emph{qualitative} study design.

% Apart from general observations, e.g., the difficulty to manage observable inconsistencies within the data structures that hold the information that is the target of tampering, e.g., the browser history and the browser cache \citep{freiling_controlled_2018}.  

In contrast to the worst case assumptions often made in the literature, in practice, adversaries do not operate under idealized circumstances. When accessing a compromised system, perpetrators usually have less control because they are under time pressure, lack knowledge of alternative methods, or need to modify the system while logged in remotely. Also, less experienced users may be able to perform actions such as changing a value in a database or editing some text in a file, but they may not be capable of booting to an alternative environment and performing low-level manipulations. 

%We denote this aspect as \emph{tamper budget}. 
In such situations, adversaries are forced to manipulate data on the system they are currently using. In contrast to dead tampering, we call this \emph{live tampering}. Live tampering is arguably not only more realistic than dead tampering, it also introduces new challenges as the act of tampering itself generates traces on the system being manipulated. We are not aware of any previous work that has specifically investigated the questions posed by live tampering.
% which the adversary must also account for while attempting to conceal their actions.

\subsection{A Qualitative Look at Live Tampering}

In this paper, we report on the results of a user study in live tampering. While prior work has shown that digital evidence tampering is hard, we focus on the question of \emph{why} this is so and therefore have chosen to apply more qualitative research methods, i.e., questionnaires and semi-structured interviews. Our general goal was to understand how study participants chose their strategies and allocated their resources while solving a live timestamp tampering task. To investigate this, we conducted a user study with 10 advanced university students, who tampered with a live system based on a fictitious scenario, in which an adversary attempts to swap two events to cover their tracks. Not all adversaries are specialist hackers or advanced persistent threats (APT) in practice, so our protagonist was assumed to be a regular user rather than a sophisticated adversary. 
As we will show, the exploration of the dynamics of such tampering also leads to understanding the difficulty of tampering with specific artifacts. This can help develop further strategies for reliable event reconstruction, since methods for representing the uncertainty of traces e.g., work by \citet{casey_standardization_2020}, include an estimate of the number of sources that agree, but also the difficulty of tampering with those sources. 

Overall, our focus was on the following research questions:
\begin{itemize}[itemsep=0ex, topsep=\parskip]

    \item[RQ1] \textbf{Approach to tampering:} What strategies do adversaries employ in planning and executing tampering with the temporal order of events?

    \item[RQ2] \textbf{Awareness and precautions of traces left by the manipulation}: How do adversaries deal with (new) traces stemming from their manipulations? 

    \item[RQ3] \textbf{Barriers to the tampering process:} What makes an artifact more difficult to tamper with compared to another?

\end{itemize}

\subsection{Contributions}

As mentioned above, it has been shown that tampering commonly results in observable inconsistencies within the data structures that hold the information that is the target of tampering. For example, manipulating browser evidence tends to result in contradictory information in the browser history and browser cache \citep{freiling_controlled_2018}. Because such tampering traces occur directly within the evidence under consideration, we call such indicators \emph{first-order} traces. With any kind of manipulation, there is always the possibility of leaving traces of the first order.

First-order traces are, however, not the only indicators of tampering that exist. Since tampering requires the use of anti-forensic tools, traces of the usage of these tools can be another type of tampering indicator. For example, timestamp manipulation tools like Timestomp can set timestamps in file systems to arbitrary values \citep{timestomp_2021}. While this eliminates traces of previous timestamps, evidence of using Timestomp usually remains. Since such traces give indirect indications of tampering, we call them \emph{second-order traces}. 

In this paper, we show that the activity of live tampering involves the consideration of second-order traces. We also exhibit tampering strategies that are opportunistic and resource-aware, i.e., where restricting factors like time pressure and lack of specific knowledge result in sub-optimal tampering strategies governed by a fixed \emph{tampering budget}.  

% From the perspective of forensic event reconstruction with its large significance for digital forensic practitioners, the insights gained by addressing these questions help to assess the confidence into an artifact and potentially find future strategies to arrive at a correct reconstruction of events despite facing contradictory temporal information.

Overall, this work provides the following contributions:
\begin{itemize}[itemsep=0ex]
    \item We present the design, implementation, and assessment of the first user study on tampering with time on a running system (live tampering). 
    \item We provide clear indications that adversaries differentiate between first-order and second-order traces and adapt their tampering strategy accordingly. 
    \item In the context of tampering with time on a live system, we identify strategies of tampering that involve the opportunistic application of tampering actions along the hierarchical abstraction layers. This indicates the mental application of a rational \emph{tampering budget}, leading to a concentration on artifacts being easier to manipulate. 
    \item Based on our analyses, we establish an understanding of the reliability of tampering indicators and derive factors that influence the tamper resistance of an artifact.
\end{itemize}

\subsection{Paper Outlook}

In the remainder of this paper, 
we first describe the details of the study itself, including the task set, the questionnaires, and the conducted interviews (Section~\ref{sec:userstudydesign}). 
%The study itself consisted of a pre/post questionnaire, selected interviews, and a tampering task.
%Sections \ref{sec:insights}, \ref{sec:results-tampering actions}, and \ref{sec:results-failures} provide a detailed analysis of the results, divided into preparation, tampering actions taken, and tampering difficulties encountered. 
The analysis of the participants' approaches and the resulting forgeries (described in Sections \ref{sec:insights}, \ref{sec:results-tampering actions}, and \ref{sec:results-failures}) reveals how they adapted their strategies based on artifact specifics as well as prior knowledge and faced difficulties, which largely revolved around (new) traces induced by their changes. 
Focusing on the technical aspects that affect tampering, we identify factors that influence the tamper resistance of artifacts allowing investigators to relatively rate their reliability. These factors are intended to serve as a basis for future developments towards the reliability of digital evidence used for event reconstruction (Section~\ref{sec:discussion}).
Overall, these findings promise to improve the reconstruction of digital forensic events and offer a new perspective on how to interpret suspected evidence tampering (Section~\ref{sec:conclusions}).

\section{User study design}
\label{sec:userstudydesign}

This section opens with a description of the tampering task scenario, followed by a detailed explanation of the user study, which consisted of four phases: (1)~a pre-tampering questionnaire, (2)~a tampering task, (3)~a post-tampering questionnaire, and (4)~a set of semi-structured interviews.

\subsection{Participants}
\label{sec:participants}
The user study was carried out within an advanced lecture on digital forensics at the Friedrich-Alexander Universität (FAU)
in the fall of 2023, with 35 registered students. 
A total of 17 students started the user study. However, only nine students worked on the task and completed the post-tampering questionnaire. Additionally, one student completed the second and third phases but did not fill out the pre-tampering questionnaire.
Given our focus on studying live tampering, only the 10 students who proceeded beyond the completion of the second phase were included in our analysis. Of these, 3 participants volunteered for the interview. 

The participants were invited to use a deterministic pseudonym to ensure a clear separation between involvement in the study and participation in the course. These pseudonyms, created in a deterministic way by combining specific non-identifying segments of personal information,  facilitated the linking of individual responses throughout the various stages of the study. In this paper, we replaced the pseudonyms to refer to them as Participants 1 to 10.

\begin{comment}
The timeline of this user study can be divided into four phases: 
\begin{enumerate}[itemsep=-1ex] 
% Potentially, compress further: topsep=\parskip
    \item[(S1)] A pre-tampering questionnaire,
    \item[(S2)] a tampering task,
    \item[(S3)] a post-tampering questionnaire, and 
    \item[(S4)] a set of semi-structured interviews.
\end{enumerate}

%\subsection{Participants}\label{sec:participants}
The user study was carried out within an advanced master lecture on digital forensics at the technical university \textbf{Blinded for review} (fall semester 2023) with 35 registered students. 
%
A total of 17 students started the user study  (completed (S1)). However, only nine students worked on the task (S2) and completed the post-tampering questionnaire (S3). Additionally, one student completed  (S2) and (S3) but did not fill out the pre-tampering questionnaire (S1) (Participant 9).
%the tampering task and the post-tampering questionnaire did not answer the pre-tampering questionnaire (Participant 9).
Given our focus on studying live tampering, only the 10 students who proceeded beyond the completion of (S1) and returned a tampered VM were included in our analysis. Of these, participants 1, 3, and 6 volunteered for the interview (S4). 
\end{comment}

\subsection{Scenario of the tampering task}
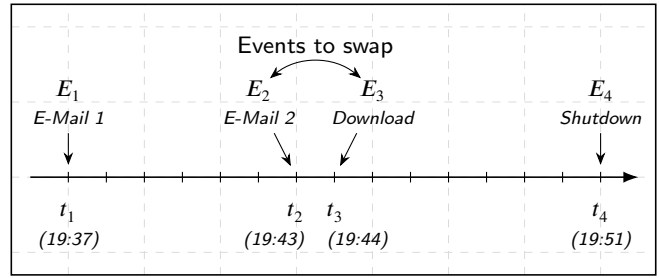
\begin{figure}[htbp]
    \centering
    %%%%%%%%%%%%%%%%%%%%%%%%%%%%%%%%%%%%%%%%%%%%%%%%%%%%%%%%%%%%%%%%%%%%%%%%
% \documentclass{standalone}
% \usepackage{tikz}

% \usetikzlibrary{backgrounds,fit, positioning, arrows, arrows.meta, calc, decorations.pathreplacing}

% \begin{document}
%%%%%%%%%%%%%%%%%%%%%%%%%%%%%%%%%%%%%%%%%%%%%%%%%%%%%%%%%%%%%%%%%%%%%%%%
\begin{tikzpicture}
\draw[help lines, color=gray!28, dashed] (-0.75,-1.3) grid (15.2*0.5,2.3);
\draw[semithick] (-.75,-1.3) rectangle (15.45*0.5,2.3);

% draw vertical lines
\foreach \x in {0,1,...,14}
\draw[] (\x*0.5,2pt) -- (\x*0.5,-2pt);
\draw [semithick,->,-Latex] (-1*0.5,0) -- (15*0.5,0);

% define coords of actions/events
\coordinate (A1) at ( 0*0.5,0) {};
\coordinate (A2) at ( 6*0.5,0) {};
\coordinate (A3) at ( 7*0.5,0) {};
\coordinate (A4) at (14*0.5,0) {};

% x axis labels
\node[align=center] at ($(A1)-(0,.65)$) {\(t_{1}\)\\\itshape\scriptsize(19:37)};
\node[align=right] at ($(A2)-(0.3,.65)$) {\(t_{2}\)\\\itshape\scriptsize(19:43)};
\node[align=left] at ($(A3)-(-0.3,.65)$) {\(t_{3}\)\\\itshape\scriptsize(19:44)};
\node[align=center] at ($(A4)-(0,.65)$){\(t_{4}\)\\\itshape\scriptsize(19:51)};

% action labels
\node[align=center] (A1Node) at ($(A1)+(0,1)$)  {\(E_{1}\)\\\itshape\scriptsize E-Mail 1};
\node[align=center] (A2Node) at ($(A2)+(-.5,1)$)  {\(E_{2}\)\\\itshape\scriptsize E-Mail 2};
\node[align=center] (A3Node) at ($(A3)+(.5,1)$)  {\(E_{3}\)\\\itshape\scriptsize Download};
\node[align=center] (A4Node) at ($(A4)+(0,1)$)  {\(E_{4}\)\\\itshape\scriptsize Shutdown};

% arrows toward axis
\draw [->,-Stealth,shorten >=0.5em] (A1Node) -- (A1);
\draw [->,-Stealth,shorten >=0.5em] (A2Node) -- (A2);
\draw [->,-Stealth,shorten >=0.5em] (A3Node) -- (A3);
\draw [->,-Stealth,shorten >=0.5em] (A4Node) -- (A4);

\draw [>=Stealth,<->] ($(A2Node.north)+(0.15,-0.1)$) to [out=40,in=140]  node[pos=0.5,above=-2pt,black, align = center, font={\small}]{Events to swap} ($(A3Node.north)+(-0.15,-0.1)$);

\end{tikzpicture}
% \end{document}
    %\vspace*{-1.5em}
    \caption{Artificial sequence of events $E_1-E_4$ imagined for the tampering task.}
    \label{fig:sequence}
\end{figure}

The user study assumes that a perpetrator who has full control over a running system (administrator/root user, such as the system owner) wants to cover the tracks of their deeds by swapping two events $E_2$ and $E_3$, as depicted in Figure~\ref{fig:sequence}, i.e., make an examiner believe that $E_3$ happened before $E_2$. 

In this synthetic scenario: 

\begin{itemize}[itemsep=0ex]
    \item $E_1$ (at time $t_1$): An email is received which asks the receiver, whether he has already seen `Rhinocerotidae', which is `really, really hot material'.
    \item $E_2$ (at time $t_2$): Another email is received in which the sender clearly states that `Rhinocerotidae' is a term referring to illegal material. %The message also contains a rhino image in an attachment. 
    The user opens the message and views the attachment.
    \item $E_3$ (at time $t_3$): The user opens a browser and issues a search query for `Rhinocerotidae', visits the Wikipedia website on `Rhinoceros', and downloads several rhino images.
    \item $E_4$ (at time $t_4$): The user shuts down the computer.
\end{itemize}

\begin{comment}
    \begin{quote}
Albert A. is accused of the illegal possession of ``rhinoceros'' images. In November 2023, the police seized his computer in his home and found several rhino images. Albert A. uses his computer at home for private purposes. Albert A. claims that he came across these images by accident, not knowing that they were illegal. In contrast, the prosecution claims that Albert A. knew that rhino images were illegal before he downloaded the images.

You are given the full computer (shut down virtual machine) of Albert A.’s computer after actions $E_1-E_4$ have been finished. 
After completing $E_3$, Albert A. thinks that sequence $E_1-E_2-E_3$ does not look good. He wants to switch actions $E_2$ and $E_3$. Play the role of Albert A. Boot the system and manipulate the computer such that “it looks as if” $E_3$ happened before $E_2$. Results will be analyzed by experts assessing the sequence of actions $E_1$, $E_2$, and $E_3$.
\end{quote}
\end{comment}

In the real world, swapping events may impact the criminal intent (`Mens rea'). 
For example, here, the updated sequence ($E_1 - E_3 - E_2$) suggests that the criminal liability of the downloaded material was unknown to the suspect when the web browsing in $E_3$ was conducted.
Although in reality, adversaries may also attempt to erase evidence of downloading illegal material, we argue that this particular type of tampering, i.e., swapping events, is a highly interesting synthetic instance that allows us to learn more than demanding trace deletion or addition. The different angles provided for the swapping of events seem to be favorable to observing the resistance and likeliness of traces and their manipulation.
% To motivate the tampering task, which is illustrated in~\cref{fig:sequence}, we argue that the artificial sequence $E_1-E_2-E_3$ will be a sign of intent and that the user knew that the downloaded images were illegal. On the contrary, the sequence $E_1-E_3-E_2$ will indicate that the user came across the images by accident and did not necessarily know that this material was subject to criminal prosecution. 

%; hence, we consider it to be a sensible experiment despite its synthetic nature.
%The swapping of events allows us to study traces as they cannot be erased and is also a highly interesting tampering instance that seems to be favorable to observing the resistance and likeliness of traces and their manipulation.

\subsection{Pre-tampering questionnaire}
The user study started with a pre-tampering questionnaire aiming at gathering a set of background information about participants. The questionnaire included a total of 15 single/multiple-choice(s) questions centered around the respondents' experience. The questions concerned their teaching curriculum such as their course of study or graduate degree (Q1-Q4, Q10, and Q11), their experience with digital forensics lectures (Q8-Q9), and practical work (Q12-Q14). 
The remaining questions aimed at capturing their workload and motivation (Q5, Q6), as well as the effort they were willing to put into the tampering task (Q7, Q15).

\subsection{Tampering task}
Participants were provided with a link to download the full disk image of a Windows 10
Home workstation (VM turned off, administrator account credentials provided) after events $E_1-E_4$ had been performed. The students were asked to act like the suspect and swap events $E_2$ (file download using Mozilla Firefox) and $E_3$ (email received via Thunderbird).\footnote{See Appendix~\ref{app:tampering_task} for the detailed instructions given to the students.} 
All actions had to be performed on the running system (working on the disk images directly was prohibited) and the time boundaries $E_1$ and $E_4$ had be to respected ($E_2, E_3$ had to remain within $E_1, E_4$).  
Otherwise, participants were free to perform any tasks such as installing arbitrary software.
%or reverting the machine through VM snapshots. %We explicitly stressed the importance of also removing traces of their manipulation, e.g., the installation of an anti-forensic product. 
%

Participants were given two weeks to complete the tampering (budget). They had to return a logbook containing notes of their actions and the VM itself.

\subsection{Post-tampering questionnaire}

Participants who returned a forged VM were asked to fill out a post-tampering questionnaire, which aimed at capturing \emph{how} the students went about the task. The questionnaire comprised 22 questions and was separated into three sections referring to a different time of the tampering task: \emph{before} (initial thoughts and strategy design), \emph{during} (tampering process), and \emph{after} (technical problems and perceived difficulties).  
Questions relating to \emph{before} and \emph{after} focused on the assessment of specific artifacts\footnote{We compiled a list of relevant, existing Win10 artifacts based on the \emph{Log2Timeline/Plaso} documentation, i.e., existing parsers which can be found in Appendix~\ref{sec:win10artifacts}.} that the respondent knew or manipulated (Q1-Q10 and Q15-Q22). For example, some questions asked participants which artifacts they manipulated and to rank them according to the perceived difficulty of tampering. The middle part targeted performed activities, e.g., how certain actions were performed (Q11-Q14). 
To answer these questions, respondents were asked to place themselves at these times and provide detailed insights into their experiences, decisions, and observations. 
They were encouraged to use their logbook to answer the questions.
%The questionnaire was developed and launched using the survey platform LimeSurvey\footnote{\url{https://www.limesurvey.org}}, which has been hosted on-premise and is available in~\ref{app:post}.

\begin{detailedversion}

In the first section of the questionnaire (before), our focus was to comprehend the students' original tampering strategies and plan of action, regarding the manipulation of specific data sources on Windows 10. The section began with two open-ended questions inviting them to detail their initial thoughts and approach to the tampering task (Q1), as well as to outline the steps they took in preparation (Q2).

For each presented artifact, respondents were asked to indicate if, at that time, they knew about it by selecting ``Yes'', ``No'', ``I am not sure'', or ``I do not remember'' (Q3). For each item of the list checked as ``Yes'', respondents were then invited to indicate if they found them relevant to solving the task (``Yes'', ``No'', ``I do not remember''), planned to manipulate them (``Yes'', ``No'', ``I do not remember''), and how hard they thought it was to manipulate them (``Easy'', ``Medium'', ``Hard'', ``I do not remember'') (Q4-Q6). The list not being exhaustive, we asked the respondents to describe any missing sources they intended to manipulate (Q7). We next asked them to rank the sources they planned to manipulate (marked as ``Yes'' in Q6) according to the perceived difficulty of tampering and the priority in which they intended to tamper with them (Q8-Q9). Lastly, we asked the respondents to give a short statement on how they specifically planned to manipulate each of these sources (Q10). 

%Although responses to these questions might be biased since they were provided after completing the tampering task, we decided not to include them in the pre-tampering questionnaire to prevent influencing participants' choices regarding the manipulation (e.g., giving them ideas about which sources they can manipulate).

The second section of the questionnaire (during) concerned the students' experiences throughout the tampering process. Through open-ended questions, we asked the respondents to describe their successes and failures (which of the planned manipulations went according to the plan, which ones failed and why - Q11), any strategy adjustments and the nature of these changes changed (Q12), any attempts to remove the effects of their activities (Q13), and any form of quality control undertaken (Q14).

In the final section (after), we asked the respondents to give a final assessment of the artifacts, based on their experiences of performing the tampering task (Q15-Q22). We repeated the structure of questions Q3-Q10 from the initial section, adjusting the tense to reflect the relevant time frame. For example, ``I knew of this source'' or ``I planned to manipulate'' were respectively changed to ``I know now of this source'' and ``I manipulated that source''.  %\end{detailedversion}

\end{detailedversion}

\begin{comment}
    \subsubsection{Misconceptions}
    We identified during the analysis of the results of part C that some participants had misconceptions about different artifacts. These can be explained as we did not provide them with a description of the sources that were listed in the questionnaire. For instance, there was confusion among some participants regarding the file system timestamps (as denoted by the \texttt{\$MFT} in the source list), labeling them as 'file internal metadata'. Others confused the \texttt{\$LogFile} system file with generic log files (e.g., Powershell history) or the Windows event logs with registry hives.
    To improve the evaluation, we corrected (when possible) these misconceptions in the following sections with contextual data originating from their logbooks or other responses. 
\end{comment}

\subsection{Interviews}
Respondents were invited to take part in semi-\-structured interviews where three individuals accepted. These interviews were conducted face-to-face and followed a generic outline with a set of open-ended questions articulated around the preliminary findings from the post-tampering questionnaire. 
The intention was to extract further insights into their experiences during the tampering task and qualitatively evaluate the difficulty of tampering with different artifacts. 
%We ran a beta test with a student from a prior user study on tampering. The comments helped us to refine the questions and estimate the duration. 
The interview comprised five areas: 
Prior experiences (experience with Windows and personal interests related to digital forensics), 
Strategy (motivations and concerns behind their decision-making, tampering approach), 
Perceived difficulty (factors that make one artifact more difficult/easier to tamper with compared to another), 
Effort (motivation and effort put into the task), and 
Conclusion (confidence and self-assessment towards the task).

\begin{detailedversion}
\begin{itemize}
    \item Prior experiences: where we asked the interviewees to evaluate their experience with the Windows operating system and to describe any personal interests related to digital forensics they had outside the teaching curriculum (e.g., participation in forensic CTF events). %reading forensic literature, or any other activities that could have contributed to their knowledge or interest in the field. 
    \item Strategy: we then invited them to reiterate their description of their tampering approach, attempting to gear the questions toward their motivations and concerns behind their decision-making
    %, concerns, and better understanding if any specific factors influenced the sequence of actions they took. 
    \item Perceived difficulty: here, we used their responses from Q20 of the post-tampering questionnaire to specifically ask the students to further elaborate on the factors that make one source more difficult/easier to tamper with compared to another. %We also invited them to describe if the perceived difficulty influenced their strategy, particularly the order of manipulation.
    \item Effort: as in the pre-tampering questionnaire, we then asked the participants to describe their motivation level and to estimate the number of working hours they put into the task.
    \item Conclusion: we ended these interviews with broader questions to evaluate their confidence and self-assessment toward the task.
\end{itemize}
\end{detailedversion}

With the written consent of the participants, the interviews were audio-recorded and lasted between 40-60 minutes. Records were later semi-automatically transcribed.

\subsection{Ethics}
The ethics commission at the university in which the user studies and interviews were conducted does not handle non-medical studies for explicit approval, and therefore instead the experimental protocol followed their general data protection and ethics rules. 

Participation in the study was voluntary and integrated into the course as a non-graded exercise. 
%
%Participation in the questionnaires did not require any personally identifiable information to be provided. Pre- and post-tampering questionnaires were provided through the university learning platform but did not require any personalized login. The student platform was also used to give general instructions, invitation links, or updates on the progress of the study. For the interviews, we invited the students to register for a slot on a Google Doc accessible without registration.  The interviewees received a % 10 
%\textbf{Value blinded for review} voucher for participating in the interviews.  
%

%Participation in the study was voluntary and integrated into the course as a non-graded exercise. 
%The data (responses, logbooks, the edited and anonymized interview transcripts) along with the questionnaires and interview guidelines are accessible here: \textbf{Blinded for review}.\footnote{Please note that, for the review process, these guidelines are included in the appendix to ensure completeness. However, they will be subsequently moved to an official repository for future reference.} 
%The raw data was stored on access-restricted servers in encrypted form and accessed only by the research team at the primary university.

\subsection{Data analysis}
We adopted an iterative approach for the analysis of qualitative data derived from the two questionnaires, logbooks, and transcriptions of the three interviews. Open-ended responses in the questionnaires and transcriptions were inductively coded to extract thematic patterns or trends directly from the respondents' answers \citep{Saldana21}. Since the logbooks primarily described the tampering process undertaken by each participant, these were not coded but used to improve our understanding of their tampering design and sequence of manipulations. 

All details to repeat the study (questionnaires and interview guidelines) are publicly available \citep{vanini2024guidelines}.

%Through careful analysis of the contextual data provided in the logbooks, we discovered prevalent misconceptions among participants regarding various artifacts. A notable instance of this was the confusion between file system timestamps (represented by the \texttt{\$MFT} in our list of artifacts) and internal metadata like EXIF metadata. This led us to make adjustments to their responses to more correctly reflect the actions they performed: Participant 1's reference to manipulating cache files was related to altering file system timestamps of cache files (change from cache files to \texttt{\$MFT}). Similarly, responses from participants who incorrectly identified their manipulation of internal metadata were revised to \texttt{\$MFT}. Additionally, Participant 7's claim of manipulating the \texttt{\$LogFile} was clarified in their logbook to be referring to the Powershell history (log file).

\subsection{Limitations}
\label{sec:limitations}
\black
The user study has three limitations: (1) it included only 10 participants which is small and may not sufficiently capture the variability of behaviors and strategies; (2) all participants were students sharing similar backgrounds and experiences; and (3) only one tampering scenario %experiment 
(switching the order of two actions) was developed which is only one possible instance of manipulation and one operationalization of the research questions, and therefore gives only answers to the research question from the viewpoint of the scenario. An adversary may have different strategies such as hiding data through file system manipulation or artifact wiping \citep{conlan_anti-forensics_2016}. 

Given these limitations, it is important to understand that our goal was a qualitative and explorative study (not a quantitative one) which allows for a detailed examination of each case. It can be seen as a first step towards better understanding strategies, challenges, and artifacts, as well as revealing new research directions such as the tampering budget, the $n$-order-trace idea, and the factors influencing the trustworthiness of timestamps. 
While some may argue that identical backgrounds are a disadvantage, our study found that participants followed different strategies which would be less informative with a large/diverse group. %Lastly, the swapping scenario involving semi-experienced adversaries reflects a common situation in digital forensic investigations which has been confirmed by practitioners.
%from the Netherlands Forensics Institute.  
%
Nevertheless, we acknowledge that a larger scale study would add value and therefore provide detailed descriptions of our tampering task which allows us to repeat this study on a larger scale. 
\black

\section{Results: Tampering Preparation}
\label{sec:insights}

This and the following two sections summarize the results we extracted from the questionnaires and interviews. We begin with insights on how participants prepared for the tampering task and what initial strategies they developed.

\subsection{Participant background, experience, and knowledge}

All participants were pursuing a master’s degree in computer science; eight were in their first year. Except for Participant 5, all had attended an introductory lecture on digital forensics. 
8 out of 9 students were moderately motivated to participate in the study. All planned to allocate 2-4 hours per week to work on the task.
Seven participants also indicated they had forensically analyzed 3-6 disk images, with 0-2 of these being on Windows systems. 

We anticipated that participants have differences in prior knowledge of (Windows) forensics which was confirmed through the post-tampering questionnaire (Q3). 
The prior knowledge level of participants can be separated into two classes: the less experienced, or novices, and the more informed participants. 
Unsurprisingly, Participants 5, 6, 8 and 9, through limited Windows Forensics experience had the lowest prior existing knowledge about the artifacts on the study system. The remaining participants had a generic knowledge of Windows artifacts. 
We noted that, with a few exceptions, participants uniformly marked all sources within the Windows Registry as well as the \texttt{\$USNjrnl} and the Thunderbird Global Database as unknown. Conversely, most participants indicated knowing Firefox-related data (cache files, cookies, and history databases), the Thunderbird's Inbox file, the \texttt{\$MFT}, the \texttt{\$RECYCLE.BIN}, and the Windows event logs. 
%These results likely mirror the topics discussed during their initial lecture on digital forensics.\todo{assumption or fact?} %Deviating from their peers, the exchange student reported solely knowing about the Firefox history, cookies databases, and Thunderbird's Inbox file.
%

During the interviews, it became evident that most participants regularly used Linux. So, even though participants had studied digital forensics, they could still be considered regular users in this scenario.%they can be seen as a regular user.  

\subsection{Participant initial thoughts and designs}
\label{initial_thoughts}
To gain a deeper insight into the participants' approaches to the tampering task, the first question of the pre-tampering questionnaire invited them to reflect on their initial thoughts and designs before commencing the task.

The responses show that all participants agreed that accomplishing the task of making it appear as if $E_3$ happened before $E_2$ required modifying one of the two events. The common strategy was therefore to select a fixed event that would act as a pivot point for re-arranging other events. 
%For instance, by choosing $E_3$ (browsing and downloading on Firefox) as their pivot, they could rearrange $E_2$ (the times of the second email) to fit the narrative. 
Interestingly, we observed that half chose $E_2$ (opening the second email) as their reference event while the other half preferred  $E_3$ (browsing and downloading on Firefox).

Participants within groups using the same pivot point expressed similar decision-making factors in their choice to re-arrange the other event: 
%their level of knowledge, their experience, the volume of linked artifacts, and the correlation with remotely stored information. %Findings are summarized in Table~\ref{tab:firefox_vs_thunderbird}. %These aspects per strategies are summarized in Table~\ref{tab:firefox_vs_thunderbird}. 

\begin{description}

\item[Level of knowledge:] Participants 4, 6, and 7 who used Thunderbird ($E_2$) as a pivot point mentioned that their decision was influenced by their greater familiarity with Firefox gained from their introductory lecture. Consequently, the task of manipulating Firefox appeared more straightforward to them.
In contrast, Participant 10 chose Firefox as the pivot point, citing limited prior knowledge and having heard that Thunderbird would be easier to manipulate.

\begin{detailedversion}
    \paragraph{Experience} Participant 5 (the exchange student) reported finding inspiration in previous tampering user studies that focused on browsing artifacts\footnote{\cite{schneider_prudent_2022}} as they had no experience with digital forensics.
\end{detailedversion}

\item[Volume of linked artifacts:] The volume of linked artifacts refers to the number of artifacts respectively connected to $E_2$ and $E_3$ that would need modification. 
%Each event is connected to a multitude of artifacts. 
For instance, notable artifacts linked to $E_2$ include Thunderbird's Inbox file, which compiles sent and received emails, and the Global database indexing their metadata. In contrast, targeting artifacts linked to $E_3$ entails manipulating a broader range of artifacts such as Firefox's history and cookies databases (\texttt{places.sqlite}, \texttt{cookies.sqlite}), cache files, and the downloaded rhino pictures.  
All participants who selected $E_3$ as their pivot point agreed that Thunderbird appeared easier to manipulate than Firefox in the post-tampering questionnaire, primarily due to the smaller volume of data involved. For four participants, the number of files to manipulate with the Firefox strategy raised concerns about increasing the likelihood of making errors. For instance, Participant 2 expressed: ``My idea was to change as little as possible to minimize the potential for errors''. For Participants 1 and 9, this also created concerns about generating too many second-order traces.
%about the effects of the manipulation i.e., leaving too many traces. 

\item[Correlation with remotely stored information:] A key factor considered by the participants was their limited control over external data storage. This concern stemmed from the possibility that an unaltered ``true copy'' of the data targeted for manipulation might exist in a remote location, like a mail server.
This concern dissuaded Participants 3, 5, and 6 of the $E_2$ pivot point group from manipulating Thunderbird.
For instance, Participant 6 commented: ``I feared that contacting the mail server could easily reveal a manipulation of the receive-timestamp''. 
%and further elaborated in the interviews that ``[...] there is somewhere a mail server which also hosts the timestamps which could pretty quickly lead for someone to see that there are some inconsistencies''. 
%
Similarly, Participant 5 indicated in their logbook that the ``full analysis of other sources than just the machine will probably give the real course of events away''. This includes sources such as ISP or connection logs, DNS request times, or Thunderbird's mail server.
Furthermore, Participant 3 shared in their interview that they had initially considered choosing Firefox as their pivot, influenced by the volume of linked artifacts factor, viewing it as a significant effort. Yet, the understanding that Thunderbird’s mail server could potentially store data remotely led them to reconsider and ultimately change their strategy. %This aspect is also reflected in further answers provided during their interview. When asked to provide explanation regarding why they had marked the OneDrive synchronization logs as ``Hard'' in (post-tampering questionnaire, Q6), the participant mentioned that the synchronization of files on external storage would be ``the same problem as with the email''.
%

% \paragraph{Correlation with the artifact logical organization} \todo{better name?} 
\item[Maintaining internal artifact consistency:]
%\sloppypar 
This factor re\-fers to the relationship between the organization of an artifact and the data it contains. For instance, SQLite databases organize their entries following a specific allocation strategy. Hence, tampering with timestamps in entries may be exposed when looking at the order of (raw) entries and identifiers in the database.
%
%A deeper understanding of the organization can reveal insights into the artifacts' authenticity or any manipulations they might have undergone. 
%
Participant 9 highlighted this issue: manipulating Firefox artifacts can be ``identified by logical inconsistencies such as timestamps not matching the order in which events are listed in a cache/log file''.

\end{description}

\subsection{Preparation}
\label{prep}
%\subsubsection{Types of preparation}
In the post-tampering questionnaire, 9 out of 10 students reported that they had undertaken preparation beforehand. We learned from their responses and their logbooks that the types of preparation steps varied from one participant to another. For example, some participants initiated the task by gathering information through a review of the literature, forums, and/or existing tools. Others adopted a `learn by doing' approach, engaging in hands-on experimentation and exploration within the VM. 
%The findings are summarized in Table~\ref{tab:prep}.

%\input{tables/preparation}

To evaluate the influence of preparation on the participants' knowledge of Windows artifacts, we compared the responses from Q3 and Q15 in the post-tampering questionnaire. 
We observed that participants we classified as informed participants in Section~\ref{sec:insights} showed minimal or no changes in knowledge levels. Conversely, most novices, i.e., Participants 5, 6, and 8, experienced a significant knowledge gain, balancing out with the overall score of the more informed participants. These participants resorted to a literature review, plus other learning resources such as testing or AI-based inquiries. 
The knowledge gained primarily concerned artifacts related to the events to modify, i.e., either Firefox or Thunderbird-related artifacts. 

Overall, the results suggest that all participants, except Participant 9, reached a comparable level of understanding, indicating that they had achieved a baseline knowledge before commencing the task. Those who reached this level of understanding before starting the task tended to initiate their work by engaging in exploration and experimentation, instead of primarily seeking information from the literature.

%
%
% \section{Insights into tampering actions}
\section{Results: Tampering Actions}
\label{sec:results-tampering actions}
This section describes the execution of the tampering task: the tampering process itself, the handling of second-order traces, as well as how quality control was performed.

\subsection{Tampering approaches}\label{sec:results-tampering actions-approaches}
All students shared the common strategy to decide on a reference event between $E_2$ and $E_3$ that is used as a pivot point to swap the other unfixed event (see Section~\ref{initial_thoughts}).  

\begin{figure*}[htbp]
% \begin{minipage}{0.48\textwidth}
\begin{subfigure}{0.48\textwidth}
  %\centering
  %%%%%%%%%%%%%%%%%%%%%%%%%%%%%%%%%%%%%%%%%%%%%%%%%%%%%%%%%%%%%%%%%%%%%%%%
% \documentclass[border=1cm]{standalone}
% \usepackage{tikz,pgfplots,pgfplotstable}
%  \pgfplotsset{compat=1.18}
% \begin{document}
% 
%
%%%%%%%%%%%%%%%%%%%%%%%%%%%%%%%%%%%%%%%%%%%%%%%%%%%%%%%%%%%%%%%%%%%%%%%%
% Data from the experiment
%
% Mapping as follows:
% 1, MFT
% 2, LNK files
% 3, Powershell history
% 4, Thunderbird Global database
% 5, Thunderbird Inbox file
%
% It's not elegant but the string mapping with pgfplots/TeX is a real pain 
%%%%%%%%%%%%%%%%%%%%%%%%%%%%%%%%%%%%%%%%%%%%%%%%%%%%%%%%%%%%%%%%%%%%%%%%

\pgfplotstableread[col sep=comma, header=true]{ 
seqnum,artifact
1,5
2,1
}\tpOne

\pgfplotstableread[col sep=comma, header=true]{ 
seqnum,artifact
1,5
2,1
3,3
4,2
}\tpTwo

\pgfplotstableread[col sep=comma, header=true]{ 
seqnum,artifact
1,5
2,1
}\tpEight

\pgfplotstableread[col sep=comma, header=true]{ 
seqnum,artifact
1,5
2,4
3,1
}\tpNine

\pgfplotstableread[col sep=comma, header=true]{ 
seqnum,artifact
1,5
}\tpTen

\begin{tikzpicture}[scale=0.95]

  \begin{axis}
    [
    height=7cm,
    xlabel={Step},
    xlabel style={font={\footnotesize\bfseries}},
    ytick={1,2,3,4,5},
    ymin=1,
    yticklabel=\empty,
    extra y ticks = {0.5,1.5,2.5,3.5,4.5,5.5},
    extra y tick labels = {
        ,% this is intentional
        \texttt{\$MFT}, 
        \texttt{LNK} files, 
        {Powershell\\history},
        {Thunderbird\\Global DB},
        {Thunderbird\\Inbox File}},
    extra y tick style={
        every tick/.style={draw=none},
        major grid style={draw=none},
        tick label style={
            font=\scriptsize, align=right
          }},
    xtick={1,2,3,4,5},
    xmin=1, xmax=5,
    xticklabel=\empty,
    extra x ticks = {1.5,2.5,3.5,4.5},
    extra x tick labels = {1,2,3,4},
    extra x tick style={
        every tick/.style={draw=none},
        major grid style={draw=none},
        tick label style={
            font=\scriptsize,
          }},
    grid=major,
    xminorgrids=false,
    yminorgrids=false,
    grid style={dashed, line width=.3pt, draw=gray!38}, 
    axis line style={semithick},
    legend style={nodes={scale=0.7, transform shape}}, 
    ]

    \addplot+[] table[x expr=\thisrow{seqnum}+0.1, y expr=\thisrow{artifact}+0.1,] {\tpOne};
    \addlegendentry{\#1}

    \addplot+[] table[x expr=\thisrow{seqnum}+0.2, y expr=\thisrow{artifact}+0.2,] {\tpTwo};
    \addlegendentry{\#2}

    \addplot+[] table[x expr=\thisrow{seqnum}+0.3, y expr=\thisrow{artifact}+0.3,] {\tpEight};
    \addlegendentry{\#8}

    \addplot+[] table[x expr=\thisrow{seqnum}+0.4, y expr=\thisrow{artifact}+0.4,] {\tpNine};
    \addlegendentry{\#9}

    \addplot+[] table[x expr=\thisrow{seqnum}+0.5, y expr=\thisrow{artifact}+0.5,] {\tpTen};
    \addlegendentry{\#10}

  \end{axis}
\end{tikzpicture}

% \end{document}
  \caption{Thunderbird group}
  \label{fig:thunderbird_seq}
\end{subfigure}
%\end{minipage}
\hfill
% \begin{minipage}{0.48\textwidth}
\begin{subfigure}{0.48\textwidth}
  % \documentclass[border=1cm]{standalone}
% \usepackage{tikz,pgfplots,pgfplotstable}
%  \pgfplotsset{compat=1.18}
%\begin{document}
%
%% Mapping of IDs to artifacts
%%
%% 1 ,$LogFile
%% 2, $USNjrnl
%% 3, $MFT
%% 4, Windows event logs
%% 5, Prefetch files
%% 6, Powershell history
%% 7, $RECYCLE.BIN
%% 8, Firefox cache files
%% 9, Firefox cookies
%% 10, Firefox history database
%
%% It's not elegant but the string mapping with pgfplots/TeX is real pain 
%
%
%% Data from the experiment
\pgfplotstableread[col sep=comma, header=true]{ 
seqnum,artifact
1,3
2,10
3,9
}\tpThree

\pgfplotstableread[col sep=comma, header=true]{ 
seqnum,artifact
1,10
2,7
3,3
4,6
5,3
}\tpFour

\pgfplotstableread[col sep=comma, header=true]{ 
seqnum,artifact
1,9
2,10
3,8
4,3
}\tpFive

\pgfplotstableread[col sep=comma, header=true]{ 
seqnum,artifact
1,10
2,9
3,3
4,5
5,6
6,2
7,1
8,7
9,4
}\tpSix

\pgfplotstableread[col sep=comma, header=true]{ 
seqnum,artifact
1,10
2,3
3,7
4,6
}\tpSeven

\begin{tikzpicture}[scale=0.95]

  \begin{axis}
    [
    height=7cm,
    xlabel={Step},
    xlabel style={font={\footnotesize\bfseries}},
    ytick={1,2,3,4,5,6,7,8,9,10},
    ymin=1, ymax=11,
    yticklabel=\empty,
    extra y ticks = {0.5,1.5,2.5,3.5,4.5,5.5,6.5,7.5,8.5,9.5,10.5},
    extra y tick labels = {
      , % this is intentional
      \texttt{\$LogFile},
      \texttt{\$USNjrnl},
      \texttt{\$MFT},
      {Windows\\event logs},
      {Prefetch\\files},
      {Powershell\\history},
      \texttt{\$RECYCLE.BIN},
      {Firefox\\cache files},
      {Firefox\\cookies},
      {Firefox\\history database},
    },
    extra y tick style={
        every tick/.style={draw=none},
        major grid style={draw=none},
        tick label style={
            font=\scriptsize, align=right
          }},
    xtick={1,2,3,4,5,6,7,8,9},
    xmin=1, xmax=10,
    xticklabel=\empty,
    extra x ticks = {1.5,2.5,3.5,4.5,5.5,6.5,7.5,8.5,9.5},
    extra x tick labels = {1,2,3,4,5,6,7,8,9},
    extra x tick style={
        every tick/.style={draw=none},
        major grid style={draw=none},
        tick label style={
            font=\scriptsize,
          }},
    grid=major,
    xminorgrids=false,
    yminorgrids=false,
    grid style={dashed, line width=.3pt, draw=gray!38}, 
    axis line style={semithick},
    legend style={nodes={scale=0.7, transform shape}}, 
    ]

    \addplot+[] table[x expr=\thisrow{seqnum}+0.2, y expr=\thisrow{artifact}+0.2,] {\tpThree};
    \addlegendentry{\#3}

    \addplot+[] table[x expr=\thisrow{seqnum}+0.3, y expr=\thisrow{artifact}+0.3,] {\tpFour};
    \addlegendentry{\#4}

    \addplot+[] table[x expr=\thisrow{seqnum}+0.4, y expr=\thisrow{artifact}+0.4,] {\tpFive};
    \addlegendentry{\#5}

    \addplot+[] table[x expr=\thisrow{seqnum}+0.5, y expr=\thisrow{artifact}+0.5,] {\tpSix};
    \addlegendentry{\#6}

    \addplot+[] table[x expr=\thisrow{seqnum}+0.6, y expr=\thisrow{artifact}+0.6,] {\tpSeven};
    \addlegendentry{\#7}

  \end{axis}
\end{tikzpicture}
\qquad 

% \end{document}
  \caption{Firefox group}
  \label{fig:firefox_seq}
 \end{subfigure}
 %\end{minipage}
    %\vspace*{-.5em}
  \caption{Sequence of manipulations. Participants are represented by lines that follow the horizontal manipulation sequence, while the vertical axis indicates manipulated objects.}
   %\vspace*{-1em}
\end{figure*}
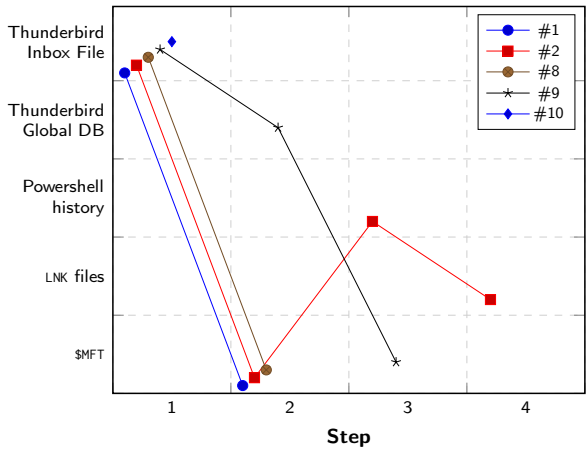
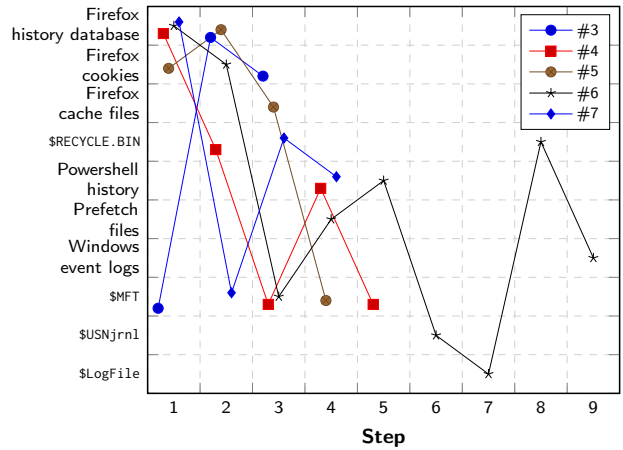

Figures~\ref{fig:thunderbird_seq} and~\ref{fig:firefox_seq} respectively illustrate the sequence of manipulations performed within the groups using the same pivot point. On the y-axis are artifacts that were manipulated and/or removed by students within each group, ordered according to their hierarchical position in the abstraction layers of the system (higher levels: application to lower levels: file system) \citep{Carrier03}. 
For example, the Firefox history database is at a higher level than the \texttt{\$USNjrnl} which belongs to the file system layer. The x-axis illustrates the sequence of steps, e.g., Participant 1 started with the Thunderbird Inbox file and further updated file system timestamps in the \texttt{\$MFT}.  

While the sequence of manipulations is also intrinsically linked to the conflicting goal of dealing with second-order traces as further discussed in Section~\ref{second_order_traces}, the graphs show a visible trend of descent through abstraction layers among participants. After choosing their pivot point, all students started by manipulating the artifacts directly related to the unfixed event to re-arrange, at the application level. For instance, participants in the Firefox pivot point group all started with the manipulation of timestamps within Thunderbird's Inbox file. We then observe that manipulations concerned artifacts in lower layers, starting with artifacts at the OS layer and addressing file system artifacts (apart from the \texttt{\$MFT}) last. %Note that Participant 3 began their manipulations by modifying the file system timestamps of Firefox cache files. 

In regards to specific timestamp manipulations, participants attempted to maintain consistency across manipulations to fit the scenario, e.g., avoiding instances where a picture's download time precedes the visit to the corresponding web page. To achieve consistency, most participants added or removed (according to the strategy) constant time offsets. Nevertheless, there were instances of less caution. For instance, one participant accidentally changed all timestamps to the same time value and others did not change all file system timestamps. Both may reveal an inconsistency in an investigation (timeline analysis).

\begin{comment}
\begin{itemize}
    \item How did they do it (timestamps): provide a new value or reduce the current value by $x$ minutes. A216 "just added 3min to each timestamp I could recognize"; A915: added a constant offset of 4min; I53: all timestamps to the same value = obvious pic in the timeline (int); N333: reduce value; R122: rewind by 4 minutes (all FS times to same value), R214: rewind by 6 minutes (pic), FS database all same value C+M, R293 reduce values.
\end{itemize}
\end{comment}

%
\subsection{Dealing with second-order traces}\label{second_order_traces}
Given the intricacies of modifying a running system, all participants showed concern about second-order traces, which is reflected in their strategy design (see Section~\ref{initial_thoughts}), choice of tooling (e.g., command prompt vs.~Powershell), and sequence of manipulations. Overall, three distinct approaches emerged from the findings. Those approaches are (1) \emph{clandestine}, (2) \emph{tampering-focused}, and (3) \emph{mixed} (discussed below).
Table~\ref{tab:tools} summarizes the tools that were used within the groups using the same approach.

\subsubsection{Clandestine approach}
Participants who used the clandestine approach proactively minimized the creation of second-order traces, rather than focusing on their removal after they had been generated. Three participants anticipated the creation of second-order traces by carefully choosing the tools they would use in the task. This included 
preferring tools that hypothetically leave fewer traces (command prompt over Powershell or Python interpreter over Notepad), 
benign-looking tools (DB Browser for SQLite),  
avoiding installing anti-forensic software \citep{conlan_anti-forensics_2016} like Timestomp \citep{timestomp_2021}, and 
using portable versions loaded on a removable medium. 
%Note that, in their interview, Participant 3 revealed a misconception by confusing Powershell with the command prompt. They used Powershell under the false belief that it would not leave any history.

Besides adjusting the manipulated files' file system metadata to align with the scenario, none of these participants attempted to eliminate further second-order traces, e.g., traces of a USB stick connection.

\subsubsection{Tampering-focused approach}
The tampering-focused approach concerns participants 5 and 8 who pertained to the act of tampering itself without considering the generation of second-order traces. The only action undertaken by both participants was the update of the file system metadata of Thunderbird's Inbox file and the transfer of files to the \texttt{\$RECYCLE.BIN}. In their logbook, Participant 5 commented that removing tools without leaving a trace seemed difficult: ``I didn't try to delete the remaining traces because I figured it would be practically impossible and would continue indefinitely''.

\begin{table}[!htb]
    \centering
    \footnotesize
        \caption{Overviews of tools to edit used by each participant during the tampering task.}
    \label{tab:tools}
    \setlength{\tabcolsep}{5pt} % Adjust the horizontal padding here
    \begin{tabular}{p{2.3cm}p{0.1cm}p{0.1cm}p{0.1cm}p{2pt}p{0.1cm}p{0.1cm}p{0.1cm}p{2pt}p{0.1cm}p{0.1cm}p{0.1cm}p{0.1cm}}
    \toprule
        &\multicolumn{3}{c}{Cland.} && \multicolumn{3}{c}{Tamp. foc.}&& \multicolumn{4}{c}{Mixed}\\
        \cmidrule(lr){2-4} \cmidrule(lr){6-8} \cmidrule(lr){10-13}
        Tools/Participants & 1 & 3 & 9 && 5 & 8 & 10 && 2 & 4 & 6 & 7\\
    \toprule
        Built-in \\
    \toprule
        cipher  & & & & & & & & & & & $\bullet$ & $\bullet$ \\
        Command prompt & & &$\bullet$ & & & & & & & $\bullet$ \\
        Powershell & & $\bullet$ & & & & & & & $\bullet$ & $\bullet$ & $\bullet$ & $\bullet$\\
        Text editor & & & $\bullet$ & & & $\bullet$ & $\bullet$ & & $\bullet$\\
    \toprule
        External\\
    \toprule
        nTimestomp & & & $\bullet$ & & $\bullet$ & & & & & & $\bullet$ \\
        Python interpreter & $\bullet$\\
        SetMace & & & & & & & &  &  $\bullet$\\
        sqlite3  & & & & & & & & & & $\bullet$ & $\bullet$ \\
        SQLite browser & & $\bullet$ & $\bullet$ &&  $\bullet$ & & & & & & & $\bullet$\\
    \bottomrule
    \end{tabular}
    %\vspace*{-1em}
\end{table}

\begin{comment}
\begin{table}[!htb]
    \centering
    \footnotesize
    \begin{tabular}{p{2.3cm}p{0.1cm}p{0.1cm}p{0.1cm}p{0.1cm}p{0.1cm}p{0.1cm}p{0.1cm}p{0.1cm}p{0.1cm}p{0.1cm}p{0.1cm}}
    \toprule
        &\multicolumn{5}{l}{Thunderbird ($n=5$)} && \multicolumn{5}{l}{Firefox ($n=5$)}\\
        \cmidrule(lr){2-6} \cmidrule(lr){8-12}
        Tools/Participants & 1& 2 & 8 & 9 & 10 && 3 & 4 & 5 & 6 & 7\\
    \toprule
        Built-in \\
    \toprule
        cipher  & & & & & & & & & & $\bullet$ & $\bullet$ \\
        CMD & & & & & & & & $\bullet$ \\
        Powershell & & $\bullet$ & & & & & $\bullet$ & $\bullet$ & & $\bullet$ & $\bullet$\\
        Python interpreter & $\bullet$\\
        Text editor & & $\bullet$ & $\bullet$ & $\bullet$ & $\bullet$\\
    \toprule
        External\\
    \toprule
        nTimestomp & & & & $\bullet$ & & & & & $\bullet$ & $\bullet$\\
        SetMace & & $\bullet$\\
        sqlite3  & & & & & & & & $\bullet$ & & $\bullet$ \\
        SQLite browser & & & & $\bullet$ & & & $\bullet$ & & $\bullet$ & & $\bullet$\\
    \bottomrule
    \end{tabular}
    \caption{Overviews of tools to edit used by each participant during the tampering task.}
    \label{tab:tools}
\end{table}
\end{comment}

\subsubsection{Mixed approach}
The mixed approach involves participants who were mindful of second-order traces generation and actively engaged in recursively removing them from the system. 
%This strategy represents a blend of the clandestine approach and what could be called a ``wrapping-the-onion" approach.
Participants in this group began by focusing on their main target (Firefox- or Thunderbird-related artifacts). They then adopted a wrapping-the-onion method, systematically working to erase not only the second-order traces of their tampering but also the subsequent layers that emerged from their efforts to conceal this tampering, and so on, in an iterative process. They also showed sophisticated efforts to minimize the generation of second-order traces by employing methods such as tool name obfuscation, downloading scripts from a homemade web server, disabling log or history monitoring, and performing the manipulation on another volume of the system. %For instance, Participant 2 

The number of tampering iterations undertaken by each participant is different. Participant 6 went the furthest as illustrated in Figure~\ref{fig:firefox_seq}.
%, by (1) removing traces of the execution of the tools used to manipulate Firefox artifacts through Powershell, (2) deleting the traces of the commands used in (1), (3) flooding the \texttt{\$USNjrnl} and \texttt{\$LogFile} to remove the traces of their changes, (4) deleting files created in (3), (5) clear Powershell's Windows Event log, and finally (6) using \texttt{cipher} to remove deleted files from the unallocated space. 
%
Interestingly, they indicated in their interview that while the perceived difficulty of tampering did not affect the order of their manipulations, the difficulty increased after each iteration, describing it as a `chain of issues', i.e., addressing one problem/trace generates a multitude of new problems/traces.

\begin{comment}
\begin{itemize}
    \item R241: "I knew that I would need tools fur updating databases and manipulating timestamps so I thought about how to remove them after usage"; I didn't want to use tools that required an installation (consisted of more than a .exe) as they leave traces that I am not aware of. + obfuscation
    \item R241: disabled history + last access timestamp update, downloaded obfuscated timestamping tools + scripts from a web server, flooding, overwriting
    \item N333: did not want to install tools, e.g., DB browser so went for command line tools (requires to download a zip anyway).
    \item R293: planned to clear the traces I left; installed using cmd to leave less trails, portable installer. + ``journal mode off''
    \item A915: Tools on USB stick; moved everything to another volume (FN) + PS/LNK files
\end{itemize}
    A216: ``I think it is very hard to have an overview of how the OS works and which traces it leaves''
\end{comment}

\subsection{Quality control}
It is reasonable to assume that some medium-skilled adversaries would perform what can be called a \emph{quality control} of their manipulations. In their logbooks and responses to question 14 of the post-tampering questionnaire, all but one participant indicated that they had conducted some form of quality control. These participants at least verified that their manipulation had been successful, sometimes with the help of artifact parsers, e.g., to read \texttt{\$USNjrnl} or \texttt{\$MFT} records. Three participants took more extensive measures, e.g., performing quality control using forensic timeline tools or tools specialized in timestomping detection. 

%Three participants took more extensive measures: Participant 9 saved the VM and then rebooted the system to verify the changes, while Participant 5 used specialized tools in timestomping detection, such as slovigz\footnote{\url{https://github.com/mariusfrinken/slogviz}} for visualizing database manipulations. Although participants were required to perform any actions from within the VM, Participant 1 loaded the disk image into Autopsy to see how the forensic tool interprets the timeline of events.  

The findings suggest that participants did not engage in overly extensive quality control measures, i.e., they did not put themselves too much in the shoes of a digital forensic examiner. From a forensic perspective, the methods used for quality control are of interest, as any activity (even verification) performed from within the VM generates additional artifacts (e.g., execution of quality control tools).

\begin{comment}
\begin{itemize}
    \item How many did it? How? $\rightarrow$ Q14 + interviews.
    \begin{itemize}
        \item I53: all same values. 
        \item Check the changes: A915 (MftRcrd), R293, T160, V318.
        \item Load into forensic tools A216 or use specialized tools (R122) slovigz (sqlite) and analyzeMFT to compare MFTs and detect timestomping. 
        \item A216: Autopsy - checked how Autopsy interprets the timeline + how Thunderbird displays the content after editing timestamps. 
        \item R241: checked only USN with forensic tools but did not have the time. 
        \item T160: saved the image and booted back up to check changes.
        \item R293 used exiftool
    \end{itemize}
    \item self-assessment $\rightarrow$ interviews $\rightarrow$ final questions 
\end{itemize}
\end{comment}

%
%
% \section{Insights into failures and difficulties in tampering}
\section{Results: Tampering Difficulties}
\label{sec:results-failures}

The majority of students experienced failures that forced them to adjust their strategy. The qualitative findings of the user study show different causes, that can be divided into two classes: (1)~practical challenges and (2)~perceptions. 
%As an example, participants reported experiencing practical issues with the application or use of tools.
%w
On top of that, most participants expressed having encountered technical difficulties when manipulating specific artifacts, which are summarized in Section~\ref{sec:factors}. 
%The results of the analysis of questions Q11 and Q12 of the post-tampering questionnaire, and the interviews show different causes. For example, a set of participants reported practical issues with the application or use of tools. Others mentioned adapting their action plan due to technical problems intrinsic to the source they attempted to manipulate. 

\subsection{Practical Challenges in Tampering} 
These aspects cover the range of issues that the participants faced in the application of the task, such as the installation or use of tools. 

When facing such practical challenges, we can see that participants were forced to make compromises on certain aspects of their strategy. For instance, Participant 3 encountered difficulties when installing Powershell modules, forcing them to install an external software and to deviate from the original plan of using only Windows' built-in tools to minimize suspicions, i.e., avoiding tools that could be considered anti-forensic tools. 
%and to reduce the number of traces that would need to be covered up.
%Participant 3 further elaborated in the interviews: ``I would have had to change the settings and I was too lazy for that and I thought afterward that SQLite Browser would not be considered a forensic tool''. 
Another participant had their external software recognized as a virus by Windows Defender and was forced to obfuscate its name and download it via a personal web server. Other participants also noted making errors when using a particular tool, such as Participant 7 who misused an overwriting tool resulting in the unintended overwrite of the entire non-allocated disk space. 

%Three participants also noted errors they made that deviated from their plan: Participants 1 and 6 respectively inadvertently neglected to manipulate log files and LNK files, while Participant 7 misused an overwriting tool, resulting in the unintended overwrite of all unallocated disk space. %Participants 

%\end{detailedversion}

%\begin{detailedversion}
\subsection{Perceptions} 
Some participants adjusted their strategy in response to non-technical factors, such as their level of knowledge and the perceived relevance of manipulating a specific artifact. For instance, Participant 6 chose not to alter the Windows event logs due to the absence of clear indications of their actions. Similarly, Participant 4 did not manipulate these logs, lacking the knowledge about where to find pertinent information and how to modify the logs. Both contemplated deleting the entire log but ultimately refrained, fearing it would appear overly suspicious. In contrast, Participant 5, unsure about how to manipulate the Firefox cache files, opted to delete the relevant ones instead. 
%
%Overall, when faced with the issue of not knowing how to manipulate an artifact, the instinctive response appears to be deletion. However, this approach seems to involve a trade-off, as the act of deletion itself could arouse suspicion.
%\end{detailedversion}

\section{Discussion and reflection}
\label{sec:discussion}
This section revisits the research questions in light of our findings. To generalize our results, we also discuss the technical aspects identified that affect the tamper resistance of specific artifacts.

\subsection{Adversary's perspective}
\label{subsec:perspective}
\textbf{RQ1: What strategies do adversaries employ in planning and executing tampering with the temporal order of events?}
When tasked with swapping two events, we observed that all participants adopted the same strategy: deciding on a reference event and using it as a pivot point to re-order the other event. 
The sequence of manipulations in re-ordering this unfixed event appeared to be closely related to the placement of each connected artifact within the hierarchical layers of the software architecture. Artifacts located at the application level were given the highest priority. As illustrated in Figures~\ref{fig:thunderbird_seq} and~\ref{fig:firefox_seq}, only a few participants manipulated traces in the lower layers along this hierarchy, which from a forensic perspective, is highly promising.
Despite planning, several participants encountered unexpected technical difficulties during their manipulation process or they reported various factors influencing their ability to tamper with certain artifacts.
Those are either inherent to the intrinsic nature of the targeted artifact, such as its complexity, or the environment in which the artifact resides (operating system, settings), e.g., the availability of tools in that environment to facilitate the manipulation. Consequently, their strategies required several adjustments while performing the tampering task. \\

\textbf{RQ2: How do adversaries deal with (new) traces stemming from their manipulations?} 
During the tampering task, participants employed three different approaches: \emph{clandestine}, \emph{tampering-focused}, and a \emph{mixed} approach. 
Of these, the mixed approach is the most sophisticated, where participants not only anticipated the creation of second-order traces but also recursively removed the traces of their deeds. In contrast, some participants considered the removal of these traces practically impossible and chose not to attempt it.
This appeared to be a never-ending conflict of goals between the manipulation of first-order traces (those associated with the event to re-order) and reducing the generation of $n+1$-order traces originating from the manipulation process itself. The participants also faced unexpected practical or technical challenges that forced them to adjust their strategy and affected their ability to create ``perfect forgeries''. In fact, as it is particularly difficult to maintain a comprehensive overview of all newly created traces on a running system, and attempting to manipulate every subsequent trace can become an endless endeavor, creating perfect forgeries might be hardly possible.

So, even if adversaries are good at covering their tracks, there may still be artifacts revealing tampering. Already with first-order traces, which are the primary focus of the manipulation, an examiner may be able to detect the tampering. For example, many tools leave discernible patterns in timestamps that may facilitate the detection of their usage \citep{galhuber_time_2021}. If the adversary is unaware of their existence or because they are constrained by a short tampering budget, they may even not consider attempting to remove them.
The detection of active tampering can also occur in every subsequent iteration of second-order, third-order traces, and so forth. In addition, tampering may generate inconsistencies in the manipulated data structures, increasing the likelihood that manipulations will be detectable.
While our tampering task focused on low/medium-skilled participants (which is probably the most common user group that local authorities investigate), the World Anti-Doping Agency v.~Russian Anti-Doping Agency case \citep{wada2020cas} has shown that even nation-state attackers leave such traces of their actions. \\

\subsection{Artifacts}

\textbf{RQ3: What makes an artifact more difficult to tamper with compared to another?} 
Our study suggests that several artifacts are more ``tamper-proof'' compared to others based on the technical challenges and difficulties faced by our tamperers:  
(1) the correlation with remotely stored information,
maintaining internal artifact consistency and the volume of linked artifacts (defined in Section~\ref{initial_thoughts} and combined here as a \emph{implicit time information} factor), as well as additional challenges that were extracted from the documentation such as
(2) the placement in the abstraction layers (see Section~\ref{sec:results-tampering actions-approaches}),
(3) the existence of integrity checks, 
(4) the assigned permissions, 
(5) encryption, and 
(6) the availability of software to edit artifacts on the system (further detailed in the next section).
These technical aspects are complemented by soft factors such as perception or knowledge which depend on the experience and sophistication of an adversary. 
%

%Interestingly, some of these factors are reflected in the mitigation recommendations of the MITRE ATT\&CK matrix for this type of manipulation.\footnote{\url{https://attack.mitre.org/techniques/T1565/001/}} \\
%%Are suggested as mitigation techniques the use of encryption, remote data storage, and restricted files and directory permissions. 

As artifacts differ in their suitability to be manipulated, this means that they have special features (or factors) that make them easy or difficult to manipulate. These factors need to be examined in more detail in the future but will be dealt with briefly in the following section.

\subsection{From tamper-proof to resistance factors} \label{sec:factors}

An adversary may (or may not) choose to tamper with an artifact for a variety of reasons, such as familiarity, perceived difficulty, or technical complexity. This section focuses on the latter and discusses aspects that prompted study participants not to make changes. The aim is to identify general factors related to artifacts that impact their resistance to manipulation. 
A more detailed discussion of this topic is presented by \citet{vanini2024evaluatingtamperresistancedigital}.
%
%Note that since these factors are based on the findings from our study, the list is not exhaustive and other factors may exist. It may also be the case that some may be merged/renamed due to dependencies or similarities: 

\begin{description}
    \item[Permissions:] Various operations on Windows are protected via User Account Control (UAC) and require Administrator privileges.
    Consequently, one factor is the level of permissions required to modify an artifact. Of course, this needs to be considered in the context of a specific user. On many system configurations, including the one in this study, the user is an Administrator of the system in question. Therefore, in many cases, the UAC interface presents little barrier to accessing the protected files, other than clicking `Allow'. On the other hand, running a command as Admin may trigger other events or be logged.
    %Some artifacts require SYSTEM privileges which needs additional tooling such as psexec or so to manipulate.

    %In a particular instance, Participant 6 indicated having experienced difficulties in tampering with Prefetch files (which are in a protected location). According to their documentation, the action of deleting Prefetch files necessitated an admin shell, which hindered their timing of actions and created further problems to cover up. When asked the question ``In your opinion, what makes a source easier to tamper with compared to another'' during the interviews, this participant further mentioned ``to have access without admin privilege requirements''. 

    \item[Integrity checks:] An artifact might have embedded mechanisms used to verify that data has not been altered or corrupted. For example, email signatures are generated over the content of the email, which may include time information. Modifying this information may result in an invalid signature and trigger suspicion. This becomes even more challenging when monitoring systems such as auditd (Linux) are used. 

    %During the tampering task, Participant 1 failed to change all timestamps related to the email from $A2$ in the Inbox file, realizing belatedly (``I never heard of it before'') that the date within the email header was part of a DKIM signature (integrated integrity check). To avoid any obvious detection due to an unverified signature, the participant decided to not tamper with this particular timestamp.\footnote{Participant 1 further elaborated in the interviews that if they had known about the signature, they would have manipulated Firefox instead.}

    % Software to edit artifacts on the system
    \item[{Software availability:}]  Manipulations require some sort of tool, which may be a text editor, regedit (both available on most Windows systems), or more specialized tools such as a hex editor or database modification tool. New tools may require an installation creating artifacts of their existence. Some may qualify as anti-forensics tools according to \cite{conlan_anti-forensics_2016} while others may be less suspicious. If no tool is available, an adversary may have to reverse engineer an artifact, which requires sophisticated knowledge. For example, utilizing knowledge of the structure of the event log file, the NSA developed the DanderSpritz module called \texttt{EventLogEdit}, which modifies integrity checks in event logs to hide manipulations.

    \item[Placement within the software stack:] This factor refers to the level at which an artifact or process is positioned within the hierarchical layers of software architecture. We believe that this impacts the modification or accessibility of an artifact, as discussed in Section~\ref{subsec:perspective}. %For example, the \texttt{\$USNjrnl} is continuously monitored and used by the operating system.

    \item[Implicit timing information:] In addition to timestamps, manipulations may lead to logical inconsistencies within an artifact. For instance, a database appends new entries at the end with an increasing ID which means potential timestamps in a column should also increase. 
    This implicit timing information may not necessarily be known to the tamperer and can now be integrated into digital forensic timelines \citep{Dreier:2024:BT}. In addition, implicit time information may also be evidence that cannot be controlled due to its residence on an external source. An example would be an Email that is sent by someone and travels through various servers.

    \item[Encryption/format:] The artifact requiring manipulation may be in a proprietary format or even encrypted (this is related to software availability) impeding a modification.  
    Considerations include the type of encryption software implementation, as well as whether the keys are available, recoverable, or not. 

\end{description}

%These general factors have an impact on the tamper resistance of artifacts. We argue that investigators should keep these factors in mind when assessing the reliability of artifacts and the likelihood of them being changed. 
% \todo{Include C-Scale? Chris - during one of our calls you had some ideas; not sure if it makes sense to put them here.}

%These factors are based on the findings from our study. We do not claim that this list is complete; other factors may occur when performing a throughout assessment of these. It may also be the case that some may be merged/renamed due to decencies or similarities. 

The possibility of evidence tampering should be considered during the investigation, encouraging examiners to look for inconsistencies. The factors presented here have an immediate impact on the tamper resistance of traces and, hence, offer a vast potential to improve the detectability of tampering because they provide clear guidance in determining the reliability of artifacts and the likelihood of them being changed. 
Given that adversaries have a finite amount of resources leading to a confined \emph{tampering budget}, they will likely fail to produce perfect forgeries. 

Detecting the act of tampering constitutes one aspect of the problem. Confronted with conflicting information, it becomes a complex endeavor for examiners to correctly reconstruct the sequence of events. 
To help investigators evaluate artifacts that contain such discrepancies, the C-Scale (`Strength of Evidence scale') can be used \citep{casey_standardization_2020}. 
It includes two core elements: the number of sources that agree \emph{and} their resistance to tampering (e.g., according to the C-Scale, the strength of evidence is higher when multiple and independent sources agree and these sources are tamper proof/more difficult to tamper with). 
Our work proposed some first tamper resistance factors which may be used to classify and interpret artifacts. 
%Ideally, this work is refined and expanded by levels. For instance, \emph{permission} may have level 1 which is user privileges, and level 2 admin privileges.

\section{Conclusion}
\label{sec:conclusions}
As outlined in the introduction, there is great interest in concealing crime. One way to do this is to tamper with digital traces afterwards. This tampering poses a significant threat to the reliability of digital investigations, especially forensic event reconstruction. Our user study sheds light on previously unexplored aspects of live system tampering.
Through a user study involving 10 graduate students tasked with swapping two past events, we identified a general tampering strategy which is to decide on one event that would act as a pivot point to re-arrange the other event.  
%a and observed the manipulation of artifacts across hierarchical system layers. 
We also concluded that manipulations generate new traces that need to be hidden or manipulated as well, resulting in an endless cycle of manipulations. Compared to dead tampering, this conflict of goals between tampering and removing the traces of tampering increases the difficulty of creating perfect forgeries. %Therefore, even if an adversary attempts to cover their tracks, there may be remaining traces 
%Consequently, a key takeaway is that it is always possible to detect tampering but it becomes more and more difficult. 
Furthermore, we generalized our results and derived factors that influence the tamper resistance of artifacts, such as embedded integrity checks and the artifact placement within the software stack. These factors are preliminary and need more discussion, however, they guide practitioners about the reliability of an artifact especially if two contradicting artifacts are found. 
We believe that the qualitative findings from our study on live tampering will improve the understanding of criminal efforts to conceal crime and aid in the interpretation as well as reconstruction of crimes.

\section*{Use of AI writing assistance}
At least one author of this paper used ChatGPT-4 and the Grammarly plugin to assist in correcting typographical and grammatical errors and refining the phrasing of certain sentences. All recommendations were thoroughly evaluated and modified when needed before being integrated into this paper.

\section*{Acknowledgments}
We wish to thank the students from the course on “Advanced Forensic Computing” at FAU for their participation. The authors also thank Linus D\"usel for the preparation of the tampering task.
This work was supported by Deutsche Forschungsgemeinschaft (DFG, German Research Foundation) as part of the Research and Training Group 2475 ``Cybercrime and Forensic Computing'' (grant number 393541319/GRK2475/2-2024). 

%%=====
%\section*{Authors contribution statement}
%Blinded for review
\printcredits{}

%%=====
\section*{Declaration of interest}
The authors declare that they have no known competing financial interests or personal relationships that could have appeared to influence the work reported in this paper.

%%       \citet{<label>} ==> Jones et al. [21]
%%       \citep{<label>} ==> [21]
%%

%% If you havbibdatabase file and want bibtex to generate the
%\bibliographystyle{elsarticle-num-names} 
\bibliography{refs.bib}

\begin{thebibliography}{27}
\expandafter\ifx\csname natexlab\endcsname\relax\def\natexlab#1{#1}\fi
\providecommand{\url}[1]{\texttt{#1}}
\providecommand{\href}[2]{#2}
\providecommand{\path}[1]{#1}
\providecommand{\DOIprefix}{doi:}
\providecommand{\ArXivprefix}{arXiv:}
\providecommand{\URLprefix}{URL: }
\providecommand{\Pubmedprefix}{pmid:}
\providecommand{\doi}[1]{\href{http://dx.doi.org/#1}{\path{#1}}}
\providecommand{\Pubmed}[1]{\href{pmid:#1}{\path{#1}}}
\providecommand{\bibinfo}[2]{#2}
\ifx\xfnm\relax \def\xfnm[#1]{\unskip,\space#1}\fi
%Type = Article
\bibitem[{Caloyannides(2003)}]{Caloyannides03}
\bibinfo{author}{Caloyannides, M.}, \bibinfo{year}{2003}.
\newblock \bibinfo{title}{Digital “evidence” and reasonable doubt}.
\newblock \bibinfo{journal}{IEEE Security \&\ Privacy} \bibinfo{volume}{1}, \bibinfo{pages}{89–91}.
\newblock \URLprefix \url{http://dx.doi.org/10.1109/msecp.2003.1266366}, \DOIprefix\doi{10.1109/msecp.2003.1266366}.
%Type = Article
\bibitem[{Carrier(2003)}]{Carrier03}
\bibinfo{author}{Carrier, B.D.}, \bibinfo{year}{2003}.
\newblock \bibinfo{title}{Defining digital forensic examination and analysis tool using abstraction layers}.
\newblock \bibinfo{journal}{Int. J. Digit. EVid.} \bibinfo{volume}{1}.
\newblock \URLprefix \url{http://www.utica.edu/academic/institutes/ecii/publications/articles/A04C3F91-AFBB-FC13-4A2E0F13203BA980.pdf}, \DOIprefix\doi{10.1109/32.588541}.
%Type = Article
\bibitem[{Casey(2020)}]{casey_standardization_2020}
\bibinfo{author}{Casey, E.}, \bibinfo{year}{2020}.
\newblock \bibinfo{title}{Standardization of forming and expressing preliminary evaluative opinions on digital evidence}.
\newblock \bibinfo{journal}{Forensic Science International: Digital Investigation} \bibinfo{volume}{32}, \bibinfo{pages}{200888}.
\newblock \URLprefix \url{https://www.sciencedirect.com/science/article/pii/S1742287619303147}, \DOIprefix\doi{10.1016/j.fsidi.2019.200888}.
%Type = Article
\bibitem[{Chisum and Turvey(2000)}]{ChisumT00}
\bibinfo{author}{Chisum, W.J.}, \bibinfo{author}{Turvey, B.E.}, \bibinfo{year}{2000}.
\newblock \bibinfo{title}{Evidence dynamics: Locard’s exchange principle \& crime reconstruction}.
\newblock \bibinfo{journal}{Journal of Behavioral Profiling} \bibinfo{volume}{1}, \bibinfo{pages}{1--15}.
%Type = Article
\bibitem[{Conlan et~al.(2016)Conlan, Baggili and Breitinger}]{conlan_anti-forensics_2016}
\bibinfo{author}{Conlan, K.}, \bibinfo{author}{Baggili, I.}, \bibinfo{author}{Breitinger, F.}, \bibinfo{year}{2016}.
\newblock \bibinfo{title}{Anti-forensics: {Furthering} digital forensic science through a new extended, granular taxonomy}.
\newblock \bibinfo{journal}{Digital Investigation} \bibinfo{volume}{18}, \bibinfo{pages}{S66--S75}.
\newblock \URLprefix \url{https://www.sciencedirect.com/science/article/pii/S1742287616300378}, \DOIprefix\doi{10.1016/j.diin.2016.04.006}.
%Type = Article
\bibitem[{Dreier et~al.(2024)Dreier, Vanini, Hargreaves, Breitinger and Freiling}]{Dreier:2024:BT}
\bibinfo{author}{Dreier, L.M.}, \bibinfo{author}{Vanini, C.}, \bibinfo{author}{Hargreaves, C.J.}, \bibinfo{author}{Breitinger, F.}, \bibinfo{author}{Freiling, F.}, \bibinfo{year}{2024}.
\newblock \bibinfo{title}{Beyond timestamps: {Integrating} implicit timing information into digital forensic timelines}.
\newblock \bibinfo{journal}{Forensic Science International: Digital Investigation} \bibinfo{volume}{49}, \bibinfo{pages}{301755}.
\newblock \URLprefix \url{https://www.sciencedirect.com/science/article/pii/S266628172400074X}, \DOIprefix\doi{https://doi.org/10.1016/j.fsidi.2024.301755}. \bibinfo{note}{dFRWS USA 2024 - Selected Papers from the 24th Annual Digital Forensics Research Conference USA}.
%Type = Article
\bibitem[{Freiling and Hösch(2018)}]{freiling_controlled_2018}
\bibinfo{author}{Freiling, F.}, \bibinfo{author}{Hösch, L.}, \bibinfo{year}{2018}.
\newblock \bibinfo{title}{Controlled experiments in digital evidence tampering}.
\newblock \bibinfo{journal}{Digital Investigation} \bibinfo{volume}{24}, \bibinfo{pages}{S83--S92}.
\newblock \URLprefix \url{https://linkinghub.elsevier.com/retrieve/pii/S1742287618300434}, \DOIprefix\doi{10.1016/j.diin.2018.01.011}.
%Type = Inproceedings
\bibitem[{Galhuber and Luh(2021)}]{galhuber_time_2021}
\bibinfo{author}{Galhuber, M.}, \bibinfo{author}{Luh, R.}, \bibinfo{year}{2021}.
\newblock \bibinfo{title}{Time for {Truth}: {Forensic} {Analysis} of {NTFS} {Timestamps}}, in: \bibinfo{booktitle}{Proceedings of the 16th {International} {Conference} on {Availability}, {Reliability} and {Security}}, \bibinfo{publisher}{Association for Computing Machinery}, \bibinfo{address}{New York, NY, USA}. pp. \bibinfo{pages}{1--10}.
\newblock \DOIprefix\doi{10.1145/3465481.3470016}.
%Type = Inproceedings
\bibitem[{Garfinkel(2007)}]{Garfinkel07}
\bibinfo{author}{Garfinkel, S.}, \bibinfo{year}{2007}.
\newblock \bibinfo{title}{Anti-forensics: Techniques, detection and countermeasures}, in: \bibinfo{booktitle}{2nd International Conference on i-Warfare and Security}, pp. \bibinfo{pages}{77--84}.
%Type = Article
\bibitem[{Harris(2006)}]{Harris06}
\bibinfo{author}{Harris, R.}, \bibinfo{year}{2006}.
\newblock \bibinfo{title}{Arriving at an anti-forensics consensus: Examining how to define and control the anti-forensics problem}.
\newblock \bibinfo{journal}{Digit. Investig.} \bibinfo{volume}{3}, \bibinfo{pages}{44--49}.
\newblock \URLprefix \url{https://doi.org/10.1016/j.diin.2006.06.005}, \DOIprefix\doi{10.1016/j.diin.2006.06.005}.
%Type = Misc
\bibitem[{Lim(2021)}]{timestomp_2021}
\bibinfo{author}{Lim, B.}, \bibinfo{year}{2021}.
\newblock \bibinfo{title}{{nTimetools}}.
\newblock \URLprefix \url{https://github.com/limbenjamin/nTimetools}.
%Type = Book
\bibitem[{Lin(2018)}]{Lin18}
\bibinfo{author}{Lin, X.}, \bibinfo{year}{2018}.
\newblock \bibinfo{title}{Introductory Computer Forensics: A Hands-on Practical Approach}.
\newblock \bibinfo{publisher}{Springer International Publishing}.
\newblock \DOIprefix\doi{10.1007/978-3-030-00581-8}.
%Type = Article
\bibitem[{Manjoo(2001)}]{Manjoo:2001:UTT}
\bibinfo{author}{Manjoo, F.}, \bibinfo{year}{2001}.
\newblock \bibinfo{title}{Unix tick tocks to a billion}.
\newblock \bibinfo{journal}{Wired} \URLprefix \url{https://www.wired.com/2001/09/unix-tick-tocks-to-a-billion/}.
%Type = Article
\bibitem[{Marrington et~al.(2011)Marrington, Baggili, Mohay and Clark}]{MARRINGTON2011S52}
\bibinfo{author}{Marrington, A.}, \bibinfo{author}{Baggili, I.}, \bibinfo{author}{Mohay, G.}, \bibinfo{author}{Clark, A.}, \bibinfo{year}{2011}.
\newblock \bibinfo{title}{Cat detect (computer activity timeline detection): A tool for detecting inconsistency in computer activity timelines}.
\newblock \bibinfo{journal}{Digital Investigation} \bibinfo{volume}{8}, \bibinfo{pages}{S52--S61}.
\newblock \URLprefix \url{https://www.sciencedirect.com/science/article/pii/S1742287611000314}, \DOIprefix\doi{https://doi.org/10.1016/j.diin.2011.05.007}. \bibinfo{note}{the Proceedings of the Eleventh Annual DFRWS Conference}.
%Type = Misc
\bibitem[{{MITRE ATT\&CK}(2020)}]{timestoming}
\bibinfo{author}{{MITRE ATT\&CK}}, \bibinfo{year}{2020}.
\newblock \bibinfo{title}{Mitre att\&ck v15.1, indicator removal: Timestomp}.
\newblock \URLprefix \url{https://attack.mitre.org/versions/v15/techniques/T1070/006/}.
%Type = Phdthesis
\bibitem[{Moch(2015)}]{Moch15}
\bibinfo{author}{Moch, C.}, \bibinfo{year}{2015}.
\newblock \bibinfo{title}{Automatisierte Erstellung von {\"{U}}bungsaufgaben in der digitalen Forensik}.
\newblock Ph.D. thesis. University of Erlangen-Nuremberg.
\newblock \URLprefix \url{https://d-nb.info/1068781181}.
%Type = Article
\bibitem[{Mohamed and Khalid(2019)}]{mohamed_detection_2019}
\bibinfo{author}{Mohamed, A.}, \bibinfo{author}{Khalid, C.}, \bibinfo{year}{2019}.
\newblock \bibinfo{title}{Detection of {Timestamps} {Tampering} in {NTFS} using {Machine} {Learning}}.
\newblock \bibinfo{journal}{Procedia Computer Science} \bibinfo{volume}{160}, \bibinfo{pages}{778--784}.
\newblock \URLprefix \url{https://www.sciencedirect.com/science/article/pii/S1877050919317119}, \DOIprefix\doi{10.1016/j.procs.2019.11.011}.
%Type = Article
\bibitem[{Neale(2023)}]{neale_fool_2023}
\bibinfo{author}{Neale, C.}, \bibinfo{year}{2023}.
\newblock \bibinfo{title}{Fool me once: {A} systematic review of techniques to authenticate digital artefacts}.
\newblock \bibinfo{journal}{Forensic Science International: Digital Investigation} \bibinfo{volume}{45}, \bibinfo{pages}{301516}.
\newblock \URLprefix \url{https://www.sciencedirect.com/science/article/pii/S2666281723000173}, \DOIprefix\doi{10.1016/j.fsidi.2023.301516}.
%Type = Article
\bibitem[{Palmbach and Breitinger(2020)}]{palmbach_artifacts_2020}
\bibinfo{author}{Palmbach, D.}, \bibinfo{author}{Breitinger, F.}, \bibinfo{year}{2020}.
\newblock \bibinfo{title}{Artifacts for {Detecting} {Timestamp} {Manipulation} in {NTFS} on {Windows} and {Their} {Reliability}}.
\newblock \bibinfo{journal}{Forensic Science International: Digital Investigation} \bibinfo{volume}{32}, \bibinfo{pages}{300920}.
\newblock \DOIprefix\doi{10.1016/j.fsidi.2020.300920}.
%Type = Book
\bibitem[{Salda{\~n}a(2021)}]{Saldana21}
\bibinfo{author}{Salda{\~n}a, J.}, \bibinfo{year}{2021}.
\newblock \bibinfo{title}{The coding manual for qualitative researchers}.
\newblock \bibinfo{publisher}{sage}.
%Type = Article
\bibitem[{Sanchirico(2004)}]{Sanchirico:2004:ET}
\bibinfo{author}{Sanchirico, C.W.}, \bibinfo{year}{2004}.
\newblock \bibinfo{title}{Evidence tampering}.
\newblock \bibinfo{journal}{Duke Law Journal} \bibinfo{volume}{53}, \bibinfo{pages}{1215--1336}.
%Type = Article
\bibitem[{Schneider et~al.(2020)Schneider, Wolf and Freiling}]{DBLP:journals/di/SchneiderWF20}
\bibinfo{author}{Schneider, J.}, \bibinfo{author}{Wolf, J.}, \bibinfo{author}{Freiling, F.C.}, \bibinfo{year}{2020}.
\newblock \bibinfo{title}{Tampering with digital evidence is hard: The case of main memory images}.
\newblock \bibinfo{journal}{Digit. Investig.} \bibinfo{volume}{32 Supplement}, \bibinfo{pages}{300924}.
\newblock \URLprefix \url{https://doi.org/10.1016/j.fsidi.2020.300924}, \DOIprefix\doi{10.1016/J.FSIDI.2020.300924}.
%Type = Techreport
\bibitem[{Strom et~al.(2020)Strom, Applebaum, Miller, Nickels, Pennington and Thomas}]{StromAMNPT20}
\bibinfo{author}{Strom, B.E.}, \bibinfo{author}{Applebaum, A.}, \bibinfo{author}{Miller, D.P.}, \bibinfo{author}{Nickels, K.C.}, \bibinfo{author}{Pennington, A.G.}, \bibinfo{author}{Thomas, C.B.}, \bibinfo{year}{2020}.
\newblock \bibinfo{title}{MITRE ATT\&CK: Design and philosophy}.
\newblock \bibinfo{type}{Technical Report} \bibinfo{number}{MP180360R1}. The MITRE Corporation.
%Type = Misc
\bibitem[{Vanini et~al.(2024a)Vanini, Gruber, Hargreaves, Benenson, Freiling and Breitinger}]{vanini2024guidelines}
\bibinfo{author}{Vanini, C.}, \bibinfo{author}{Gruber, J.}, \bibinfo{author}{Hargreaves, C.}, \bibinfo{author}{Benenson, Z.}, \bibinfo{author}{Freiling, F.}, \bibinfo{author}{Breitinger, F.}, \bibinfo{year}{2024}a.
\newblock \bibinfo{title}{Guidelines and questionnaires for a user study on live timestamp tampering in digital forensic event reconstruction}.
\newblock \DOIprefix\doi{https://doi.org/10.48657/ydrk-qa98}. \bibinfo{note}{(Version 1.0)}.
%Type = Misc
\bibitem[{Vanini et~al.(2024b)Vanini, Hargreaves and Breitinger}]{vanini2024evaluatingtamperresistancedigital}
\bibinfo{author}{Vanini, C.}, \bibinfo{author}{Hargreaves, C.}, \bibinfo{author}{Breitinger, F.}, \bibinfo{year}{2024}b.
\newblock \bibinfo{title}{Evaluating tamper resistance of digital forensic artifacts during event reconstruction}.
\newblock \DOIprefix\doi{10.48550/arXiv.2412.12814}, \href{http://arxiv.org/abs/2412.12814}{\tt arXiv:2412.12814}.
%Type = Inproceedings
\bibitem[{Willassen(2008)}]{Willassen2008}
\bibinfo{author}{Willassen, S.Y.}, \bibinfo{year}{2008}.
\newblock \bibinfo{title}{Finding evidence of antedating in digital investigations}, in: \bibinfo{booktitle}{Proceedings of the 2008 Third International Conference on Availability, Reliability and Security}, \bibinfo{publisher}{IEEE Computer Society}, \bibinfo{address}{USA}. p. \bibinfo{pages}{26–32}.
\newblock \URLprefix \url{https://doi.org/10.1109/ARES.2008.149}, \DOIprefix\doi{10.1109/ARES.2008.149}.
%Type = Misc
\bibitem[{{World Anti-Doping Agency}(2020)}]{wada2020cas}
\bibinfo{author}{{World Anti-Doping Agency}}, \bibinfo{year}{2020}.
\newblock \bibinfo{title}{Cas 2020/o/6689 world anti-doping agency v. russian anti-doping agency}.
\newblock \bibinfo{howpublished}{\url{https://www.tas-cas.org/fileadmin/user_upload/CAS_Award_6689.pdf}}.

\end{thebibliography}

%\begin{thebibliography}{00}

%
%\bibitem[Author(year)]{label}
%\bibliographystyle{elsarticle-num-names} 
%\bibliography{references.bib}



%\end{thebibliography}

%\clearpage
\appendix

\section{Tampering task}
\label{app:tampering_task}
%Participants were given full access to a shut-down VM, embodying Albert A.'s ``full computer'' after actions $E_1-E_4$ had been performed.
%
%The VM was made available via a download link and participants received the password of the session during the task presentation. After receiving the VM, the participants were asked to continue the fictitious scenario:
The following case scenario and task were given to the students: 

\begin{quote}
Albert A. is accused of the illegal possession of ``rhinoceros'' images. In November 2023, the police seized his computer in his home and found several rhino images. Albert A. uses his computer at home for private purposes. Albert A. claims that he came across these images by accident, not knowing that they were illegal. In contrast, the prosecution claims that Albert A. knew that rhino images were illegal before he downloaded the images.

You are given the full computer (shut down virtual machine) of Albert A.’s computer after actions $E_1-E_4$ have been finished. 
After completing $E_3$, Albert A. thinks that sequence $E_1-E_2-E_3$ does not look good. He wants to switch actions $E_2$ and $E_3$. Play the role of Albert A. Boot the system and manipulate the computer such that “it looks as if” $E_3$ happened before $E_2$. Results will be analyzed by experts assessing the sequence of actions $E_1$, $E_2$, and $E_3$.
\end{quote}

\section{Windows 10 forensic artifacts}
\label{sec:win10artifacts}

Before designing the questionnaire, we compiled a list of relevant, existing Windows 10 artifacts based on the \emph{Log2Timeline/Plaso} documentation.
This included artifacts within the registry and other user application or OS-related artifacts such as LNK files or the \texttt{\$RECYCLE.BIN}. In addition, we added several artifacts that we deemed relevant but are currently not considered by Plaso such as Thunderbird's Inbox file and Global database (email index system). The complete list can be found below in Table~\ref{tab:sources_timing_info}.

%%%%%%%%%%%%%%%%%%%%%%%%%%%%%%%%%%%%%%%%%%%%%%%%%%%%%%%%%%%%%%%%%%%%%%%%
%%%%%%%%%%%%%%%%%%%%%%%%%%%%%%%%%%%%%%%%%%%%%%%%%%%%%%%%%%%%%%%%%%%%%%%%
\begin{comment}
\begin{table*}
    \centering
    \footnotesize
    \begin{tabular}{p{6cm}p{6.4cm}p{4cm}}
    \toprule
         Application & OS &File system \\
    \midrule
         Files internal metadata &Amcache (registry) & \texttt{\$LogFile}\\
         Firefox cache files &Bam (registry)&\texttt{\$MFT} \\
         Firefox cookies &Jumplists  &\texttt{\$RECYCLE.BIN}\\
         Firefox history and downloads database &LNK files &\texttt{\$USNjrnl}\\
         Microsoft Edge cache files  &OpenSavePIDMRU / LastVisitedPIDMRU (registry)\\
         Microsoft Edge history and downloads database &Prefetch files  \\
         OneDrive synchronization logs &\texttt{setupapi.dev.log} \\
         Thunderbird Inbox file &Shellbags (registry)\\
         Thunderbird Global database &ShimCache (registry)\\
         &USB/USBSTOR (registry) \\
         &UserAssist (registry) \\
         &Windows Event Logs \\
         &Windows timeline database\\
    \bottomrule
    \end{tabular}
    \caption{Catalog of Windows artifacts derived from Plaso parsers.}
    \label{tab:sources_timing_info}
\end{table*}
\end{comment}

\begin{table}[!b]
    \centering
    \scriptsize
    \caption{Catalog of Windows artifacts derived from Plaso parsers.}
    \begin{tabular}{p{1.5cm}p{5.5cm}}
    \toprule
         Layers & Sources \\
    \midrule
        Application &Files internal metadata\\
         &Firefox cache files \\
         &Firefox cookies \\
         &Firefox history and downloads database \\
         &Microsoft Edge cache files \\
         &Microsoft Edge history and downloads database \\
         &OneDrive synchronization logs \\
         &Thunderbird Inbox file \\
         &Thunderbird Global database \\
    \midrule
         OS &Amcache (registry)  \\
         &Bam (registry) \\
         &Jumplists \\
         &LNK files \\
         &OpenSavePIDMRU / LastVisitedPIDMRU (registry) \\
         &Prefetch files \\
         &\texttt{setupapi.dev.log}\\
         &Shellbags (registry) \\
         &ShimCache (registry) \\
         &USB/USBSTOR (registry) \\
         &UserAssist (registry) \\
         &Windows Event Logs \\
         &Windows timeline database\\
    \midrule
         File system & \texttt{\$LogFile}\\
          &\texttt{\$MFT}  \\
         &\texttt{\$RECYCLE.BIN} \\
         &\texttt{\$USNjrnl} \\
    \bottomrule
    \end{tabular}
    \label{tab:sources_timing_info}
\end{table}

\begin{comment}
\begin{table}
    \centering
    \footnotesize
    \begin{tabular}{p{6.4cm}}
    \toprule
         Sources \\
    \midrule
         \texttt{\$LogFile} \\
         \texttt{\$MFT}  \\
         \texttt{\$RECYCLE.BIN} \\
         \texttt{\$USNjrnl} \\
         Amcache (registry)  \\
         Bam (registry) \\
         Files internal metadata\\
         Firefox cache files \\
         Firefox cookies \\
         Firefox history and downloads database \\
         Jumplists \\
         LNK files \\
         Microsoft Edge cache files \\
         Microsoft Edge history and downloads database \\
         OneDrive synchronization logs \\
         OpenSavePIDMRU / LastVisitedPIDMRU (registry) \\
         Prefetch files \\
         \texttt{setupapi.dev.log}\\
         Shellbags (registry) \\
         ShimCache (registry) \\
         USB/USBSTOR (registry) \\
         UserAssist (registry) \\
         Thunderbird Inbox file \\
         Thunderbird Global database \\
         Windows Event Logs \\
         Windows timeline database\\
    \bottomrule
    \end{tabular}
    \caption{Catalog of Windows artifacts derived from Plaso parsers.}
    \label{tab:sources_timing_info}
\end{table}
\end{comment}

%\section{}

\end{document}